\documentclass[12pt]{article}
\usepackage{epsfig}
\usepackage{graphicx}

\usepackage{amssymb,amsmath}

\def\la{\langle }
\def\ra{ \rangle }
\def\la{\langle }

\newcommand{\beq}{\begin{equation}}
\newcommand{\eeq}{\end{equation}}
\newcommand{\bea}{\begin{eqnarray}}
\newcommand{\eea}{\end{eqnarray}}

\newcommand{\DL}{\mathcal{D}}
\newcommand{\PL}{\mathcal{P}}

\newcommand{\jpmdisclaimer}{\vskip3cm\par\noindent\textbf{\footnotesize
Information has been obtained from sources believed to be reliable
but JPMorgan Chase \& Co. or its affiliates and/or subsidiaries
(collectively JPMorgan) does not warrant its completeness or
accuracy.  Opinions and estimates constitute our judgement as of the
date of this material and are subject to change without notice.
Past performance is not indicative of future results.  This material
is not intended as an offer or solicitation for the purchase or sale
of any financial instrument.  Securities, financial instruments or
strategies mentioned herein may not be suitable for all investors.
The opinions and recommendations herein do not take into account
individual client circumstances, objectives, or needs and are not
intended as recommendations of particular securities, financial
instruments or strategies to particular clients. The recipient of
this presentation must make its own independent decisions regarding
any securities or financial instruments mentioned herein.  JPMorgan
may act as market maker or trade on a principal basis, or have
undertaken or may undertake account transactions in the financial
instruments or related instruments of any issuer discussed herein
and may act as underwriter, placement agent, advisor or lender to
such issuer. JPMorgan and/or its employees may hold a position in
any securities or financial instruments mentioned herein.  Important
disclosures are available on a company specific basis at
http://mm.jpmorgan.com/disclosures/company.  Additional information
available upon request.}}

\begin{document}
\begin{titlepage}
\begin{flushright}
\end{flushright}
\vskip3cm
\begin{center}
{\LARGE
Implied Multi-Factor Model for Bespoke CDO Tranches
\vskip0.5cm
and other Portfolio Credit Derivatives}
\vskip1.0cm
{\Large Igor Halperin}
\vskip0.5cm
Quantitative Research, JP Morgan
\vskip0.5cm
Email: igor.halperin@jpmorgan.com
\vskip0.5cm
October 2009 \\

\vskip1.0cm
{\Large Abstract:\\}
\end{center}
\parbox[t]{\textwidth}{
This paper introduces a new semi-parametric approach to the 
pricing and risk management of bespoke CDO tranches, with a particular attention to 
bespokes that need to be mapped onto more than one reference portfolio.
The only user input in our framework is a multi-factor model (a "prior" model 
hereafter) for index portfolios, such as 
CDX.NA.IG or iTraxx Europe, that are chosen as benchmark securities for the pricing 
of a given bespoke CDO. Parameters of the prior model are fixed, and not tuned to match
prices of benchmark index tranches.
Instead, our calibration procedure amounts to a proper reweightening 
of the prior measure using the Minimum Cross Entropy method. As the 
latter problem reduces to convex optimization in a low dimensional space, our model is
computationally efficient.
Both the 
static (one-period) and dynamic versions of the model are presented. The latter 
can be used for pricing and risk management of more exotic instruments referencing 
bespoke portfolios, such as forward-starting tranches or tranche options, and
for calculation of credit valuation adjustment (CVA) for bespoke tranches.  
}

\vspace{2.0cm}
\newcounter{helpfootnote}
\setcounter{helpfootnote}{\thefootnote}
\renewcommand{\thefootnote}{\fnsymbol{footnote}}
\setcounter{footnote}{0} \footnotetext{ Opinions expressed in this
paper are those of the author, and do not  necessarily reflect the
view of JP Morgan. I would like to thank Andrew Abrahams, Anil Bangia and Jakob Sidenius for valuable discussions.
All remaining errors are my own.
}
\renewcommand{\thefootnote}{\arabic{footnote}}
\setcounter{footnote}{\thehelpfootnote}

 \end{titlepage}
 
\section{Introduction}

It is well known that virtually all models for pricing and risk management of derivative 
securities are interpolators of different degrees of sophistication. Given observed 
market prices of various derivatives, the problem of pricing a new, illiquid security, whose price
is not available from the marketplace, amounts to first finding benchmark 
traded  (liquid) derivatives most 
similar to the 
one at hand, and then providing a rule by which their prices should be interpolated (and possibly 
extrapolated) in order to price our security.

The second step is where one needs a model. Complexity of the model depends on the instrument one
needs to price, and on the market. In some cases, 
the model can be as simple as a one-dimensional spline (or even linear) interpolation. This 
is the case when e.g. an equity option with a given strike and maturity is priced by 
interpolating implied 
volatilities of benchmark traded options on the same underlying and with the same maturity.
Note that generally such interpolation is done not directly in 
the price space, but in some "model parameter space", a procedure that 
should properly handle constraints imposed by the no-arbitrage principle, and also 
account for stochasticity of 
underlying price processes.
In particular, 
as long as the underlying for both the illiquid and liquid options is the same, pricing by
interpolation is relatively straightforward once one specifies a stochastic model for the 
underlying, calibrates it, and identifies relevant parameters that 
differentiate between the benchmark and the target derivatives. 
 
In the case of correlation-dependent portfolio credit derivatives, the role of benchmark 
securities is played
by standardized tranches referencing standard portfolios, called credit index portfolios, 
such as CDX.NA.IG or iTraxx Europe. Both these portfolios consist of 
synthetic exposures (in the credit default swap format) to 125   
liquidly traded investment grade corporate obligors (``names''), added 
with equal weights. Index tranches are swap contracts
covering portfolio losses between an attachment and detachment points (commonly 
referred to as 
strikes) that are expressed as a percentage of the total portfolio notional.
Standardized strikes are 
0, 3, 7, 10, 15, 30, 100 \% for CDX.NA.IG index, and 0, 3, 6, 9, 12, 22, 100 \% for iTraxx 
Europe index, and 
standard tradeable maturities are 5, 7, 10 and (to a lesser extend) 3 years. Other standard
portfolios are somewhat less liquid, but still are more or less actively traded in the market.
In particular, CDX.NA.HY index has 100 US non-investment grade (``high yield'') names, with 
standardized strikes of 0, 10, 15, 25, 35 and 100 \%.   

Market participants use quoted prices of standardized index tranches in order to estimate
prices of other correlation-dependent derivatives, such as cash CDO tranches, non-standard
tranches referencing standard index portfolios with strikes or/and maturities different from
those actively traded, or customized (bespoke) synthetic tranches whose reference portfolios
differ in their composition from credit index portfolios. More specifically, as prices of 
these 
instruments are driven by both observable (CDS prices) and unobservable parameters (in 
particular, parameters determining default dependencies under the risk-neutral 
measure, such as asset correlations in the Gaussian copula model), practitioners 
typically use market prices
to estimate the latter set of parameters using a specific model, and then use (and possibly 
interpolate/extrapolate) these parameters
within the same model to price the instrument in question. 

In this paper, we address the problem of pricing sythnetic bespoke CDO tranches by
unraveling of information contained in the market prices of benchmark index tranches.
In the Street jargon, this is often referred to as ``bespoke mapping problem'', implying 
that a given bespoke tranche is ``mapped'' onto index tranches.
Obviously, this problem can be viewed as another example of pricing by interpolation.
On the other hand,
it is clear that, as long as composition of a bespoke portfolio is different from that 
of an index portfolio, the  bespoke mapping problem is considerably more complicated than 
the above problem of pricing an illiquid equity option by interpolation of Black-Scholes 
implied volatilities.
The reason is that the present case  requires not only  
interpolation  across strikes and/or maturities, but should also somehow involve 
interpolation/extrapolation across the underlying.

As the Gaussian copula with base correlations (and a possible extension to 
random recovery) continues,
in spite of its well known drawbacks, to serve as the current market standard, 
practitioners usually
pose the bespoke mapping problem as the base correlation mapping problem. The idea here is 
to find proper correlation parameters for a bespoke tranche at hand by a suitable 
adjustment of 
market-implied base correlations for standard index tranches, with an adjustment designed to 
account for differences between the index and bespoke portfolios.

While various base correlation mapping rules 
are used by practitioners, these rules are {\it ad hoc} and lack a theoretical 
or empirical justification. Worse yet, the base correlation method  
is theoretically inconsistent, and occasionally violates no-arbitrate constraints in practice, 
even when applied to a simpler problem of pricing non-standard tranches referencing the same 
index portfolio. Furthermore,  being a one-factor model, the base correlation approach does not properly 
addresses the sector
concentration risk, and thus cannot be expected to provide thustworthy results as long as 
composition of a bespoke portfolio is materially different from that of the reference index 
portfolio. More details on the base correlation mapping rules and their drawbacks will be given below. 

In this paper we develop a consistent and practically oriented model for pricing despoke 
CDOs and
other portfolio credit derivatives.
We specifically concentrate on the case of bespokes that 
have to be mapped onto more than one reference index, which is often the case in practical
settings, where a given bespoke portfolio can include e.g. both US and European names, or both 
investment and non-investment grade (IG and HY, respectively) names. Comparing to bespokes that
need only be mapped onto one reference index, the latter case is more complex and produces higher modeling uncertainty. At the same time, it is also more computationally
demanding, as it calls for a multi-factor framework (see below) in which one has to simultaneously
calibrate to tranches written on all reference indices.

Our framework combines several modeling concepts, and generally belongs in the 
class of ``implied distribution'' models that have gained in popularity in recent years. We 
start with a bottom-up view of an index portfolio within a factor framework with a 
multivariate 
``market factor'' $ \vec{Z} $, where individual defaults
are independent conditional on the value of $ \vec{Z} $. Any arbitrage-free
factor model of conditional
independence type can be used here, examples are discussed below. This model is referred to as the {\it prior} model. At the next step, we departure from the usual 
bottom-up framework in two aspects. First, we assume that parameters in our prior
model are fixed once and for all, i.e. we do {\it not} fit this model to available 
data. Second, we give up the fine resolution of the portfolio loss   
into contributions of individual obligors, and instead
 bucket all names  in the index portfolio (and their respective losses) into two groups of 
 names (sub-portfolios) according 
to their membership degree in the bespoke portfolio\footnote{I.e. 
the first group includes all names from the index portfolio that enter the 
bespoke portfolio, and the second group includes all the rest.}. 
In this way we construct a 
{\it dichotomic representation} of index portfolio losses as a joint loss distribution 
of two sub-portfolios. As an avatar of the factor framework we have started with, losses 
in the  
sub-portfolios are independent conditional on the market 
factor. 
This representation serves  
as a single object of further analysis. 

In the second step, we 
calibrate our model by finding a minimal functional distortion of the prior dichotomic
loss distributions of index portfolios
needed   
in order to match observed  prices of tranches referencing these portfolios.
This is done within a semi-parametric framework based on the Minimum Cross Entropy (MCE)
method, where the number of free parameters 
is data-driven, and is equal to the number of tradable tranches.
The resulting joint loss 
distribution is then used to price tranches on a bespoke portfolio. 
Details of this procedure will be introduced in Sect.3.


The justification for taking a top-down view of the portfolio loss 
process is three-fold. First, market 
incompleteness imposes severe restrictions on the extent to which the
"true" (market-implied) risk-neutral measure can be learned. For example,
for a standard
portfolio like CDX.NA.IG or iTraxx Europe, the only available source of information about 
risk-neutral dependencies are tranche prices (6 quotes per maturity), which is not 
sufficient to infer the pricing measure in a unique way.
The only way out is to model it using a low dimensional space of adjustable parameters.
With our MCE approach, such a parametrization is constructed directly in the loss
space, no-arbitrage relations are satisfied by construction, and calibration 
amounts to convex optimization\footnote{Note that calibration of most of bottom-up 
models amounts to non-convex optimization with multiple local minima, which in practice 
often leads to unstable calibration and hedging.}. Second, 
as will be discussed in more details in Sect 2.4, the knowledge of
dichotomic loss distributions of index portfolios is sufficient to 
price tranches on bespoke portfolios made of arbitrary compositions  
of "chunks" of index portfolios, and can be readily generalized to other types of bespoke
portfolios which involve names not belonging to any index.
Third, our approach is computationally more efficient than a  
bottom-up one which can become quite computationally intense in a typical
setting of pricing bespoke tranches that need be mapped onto more than one index 
portfolio.

The rest of this paper is organized as follows. The reminder of this introduction 
discusses base correlation mapping methods and explains the relation of our approach to 
the previous literature.  
In Sect.2 we provide a high-level qualitative overview of key elements  
of our modeling framework. Sect.3 provides technical details of the 
Minimum Cross Entropy (MCE) calibration scheme.
Information-theoretic aspects of our problem are discussed in the appendix. 
In Sect.4 we present a dynamic version of the model. 
Sect.5 deals with generalization and extensions. Numerical examples are considered in Sect.6.
The final Sect.7 concludes.        
      

\subsection{Base correlation mapping methods}

A one-factor Gaussian copula model developed by Li \cite{Li} continues to be the market 
standard
model. As in its original form the model is uncapable of  matching market prices, practitioners use 
it in a way similar to the way the Black-Scholes model is used in other markets. 
Namely, traders
convert market prices of index tranches into what is called base correlations. The latter 
are 
defined in two steps. First, one converts market prices into prices of synthetic equity 
tranches 
(base tranches) covering losses from 0 \% to $ K \% $, where $ K $ is one of the standard 
strikes for 
the index under consideration. Second, base correlation $ \beta(K)$ for strike $ K $ is 
found as 
the correlation parameter that should be used by the Gaussian copula model in order to 
match the 
price of base tranche with strike $ K $. Market prices typically imply an increasing 
``correlation skew'' function $ \beta(K) $.

Within the Gaussian copula/base correlation framework, the problem of pricing bespoke 
tranches
amounts to calculating base correlations for a bespoke portfolio from base 
correlations for
an index portfolio. This is achieved by first calculating strike $ K_i $ of the index 
portfolio that is 
``equivalent'' (in a sense defined below) to strike $ K_b $ of the bespoke tranche, and 
then using interpolation of 
base correlations for standard strikes in order to find $ \beta(K_i) $.  
Several methods are used by traders to define what is meant by such ``equivalence'':
\begin{itemize}
\item Absolute strike rule (no mapping): Here to price a bespoke tranche with
strike $ K_b $, we use the base correlation for the same strike, i.e. our rule is 
$ K_i = K_b $. Such a naive rule is seldom used as it stands, as the probabilities to 
reach the same strike for a bespoke and index portfolio can be very different
if e.g. their 
average spreads are vastly different.
\item ATM mapping rule: Here one assumes that the observed correlation 
skew $ \beta(t, K,L,\ldots)$ (written here as a general function of calendar time $ t $, 
strike $ K $, expected portfolio loss $ L $ and possible other variables) is 
a function of 
$ t  $ and a single
dimensionless ratio (``moneyness'') $ x \equiv K/L $:
\beq
\label{ansatz}
\beta(t,K, L,\ldots) = \beta \left(t, x = \frac{K}{L} \right)
\eeq
(Note that if the skew is a function of {\it only} $ t, K $ and $ L $, then the ansatz
(\ref{ansatz}) follows simply on the dimensional grounds.)
  
Given expected losses $ L_b $ and $ L_i $ of the bespoke and 
index portfolios, respectively, the two portfolio will have the same $ \beta $, and 
hence the same risk, when they have the same value of $ x  $. This yields 
the following relation for the index strike $ K_i $ 
that is "equivalent" to a given strike $ K_b $ of the bespoke portfolio:
\beq
\label{ATM}
\frac{K_i}{L_i} = \frac{K_b}{L_b}
\eeq
It is interesting to note that ansatz (\ref{ansatz}) has some {\it dynamic} 
implications for the behavior 
of the correlation skew. Indeed, let us calculate its partial derivatives:
\bea
\label{partial}
\frac{\partial \beta}{\partial K} &=& \frac{1}{L} \frac{d \beta}{d x} \\
\frac{\partial \beta}{\partial L} &=& - \frac{K}{ L^2} \frac{ d \beta}{d x} = 
- \frac{K}{L} \frac{\partial \beta}{\partial K} \nonumber
\eea
If the skew is upward sloping, then the second of 
Eqs.(\ref{partial}) implies that $ \partial \beta / \partial L \leq 0 $, i.e. the skew moves
downward as the expected portfolio loss (and hence the par portfolio spread) increases.

If we assume that the correlation skew $ \beta(t,x) $ is a slowly changing function 
of its first argument\footnote{Equivalently, we assume that the characteristic scale 
on which $ \beta(t,x) $ changes as a function of its 
first argument is of order of the length of an economic cycle, i.e. a few years.
Note that this is the same assumption that is tacitly made anyway in using single-period 
models such 
as CreditMetrics/Gaussian copula model for pricing CDO tranches.
}, i.e. its changes are largely driven by changes of $ x $,  then
the latter prediction can be tested using the actual data. In
Fig.~\ref{fig:skew_CDX_NA} we show the profile of the correlation skew of 
the CDX.NA.IG11 index portfolio for 04/15/09 and 05/15/09. One sees that while
the index spread goes down, the skew moves down as well. 
We conclude that the ATM rule (\ref{ATM}) is likely to be in conflict with empirical data.
\begin{figure}[t]
\begin{center}
\includegraphics[width=10cm, height=60mm]{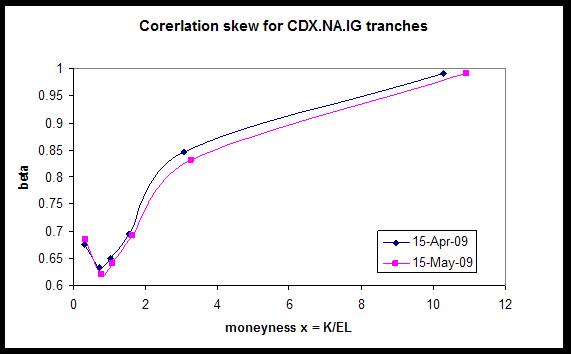}
\caption{Correlation skew for CDX.NA.IG11 as a function of moneyness.
Expected portfolio loss is 9.72 \% on 04/15/09 and 9.16 \% on 05/15/09.} 
\label{fig:skew_CDX_NA}
\end{center}
\end{figure}

What is missing in Eq.(\ref{ATM})? Note that the strike adjustment according to 
(\ref{ATM}) is determined only by the portfolio expected loss and not e.g. higher moments
of the loss distribution. This may appear counter-intuitive.  Indeed, comparing two 
portfolios having e.g. the same average spread but different spread dispersion, we 
expect that the more dispersed portfolio 
should have higher risk under nearly any reasonable 
risk metric\footnote{In terms of the dimension analysis , we can expect that, in 
addition 
to $ x = K/ \la L \ra $, correlation can depend 
on $ K/ \sqrt{\la L^2 \ra - \la L \ra^2} $ and other dimensionless
combinations.}.
 
\item Probability matching rule: $ K_i $ is determined from the condition that 
cumulative probability
of losses reaching strikes $ K_b $ and $ K_i $ for the bespoke and index portfolios should match:
\beq
\label{ProbMatch}
Pr \left( L_i \leq K_i \right) = Pr \left( L_b \leq K_b \right) 
\eeq
This prescription takes higher moments of the loss distributions for the bespoke and 
index portfolio into account. However, the problem now is that the right-hand side (RHS)
of (\ref{ProbMatch}) is only defined as long as the loss distribution for the bespoke 
is known, but the latter is exactly what we want to find in the first place!
This means that on its own, Eq.(\ref{ProbMatch}) is incomplete, and thus potentially has 
an infinite number of solutions.
The way the rule (\ref{ProbMatch}) is used in practice is to {\it assume} that 
its RHS can be computed using CreditMetrics/Base Correlation model with correlation
$ \beta( K_i ) $. When interpreted in this way, Eq.(\ref{ProbMatch}) becomes 
a highly non-linear equation for $ K_i $, whose fixed point determines both the 
equivalent strike and the pricing measure for the bespoke.
However, it is important to realize that this assumption is {\it ad-hoc}, and may lead 
to a sub-optimal and/or biased solution of the bespoke mapping problem\footnote{More 
specifically, we assert that the bespoke pricing measure obtained with this prescription
{\it is} sub-optimal according to the information-theoretic MCE method which 
allows to chooce a ``best'' bespoke measure among possible solutions of 
(\ref{ProbMatch}), see Sect.3.}.  
As a practical problem, the probability matching method does not always have a solution for 
bespokes with wide spreads.
\end{itemize}

While other similar recipes of base correlation mapping are available in 
the literature, see e.g. \cite{SocGen}, 
the main problem with all these methods is that they are {\it ad hoc} and do not 
have a solid theoretical or empirical basis. In practice, their use leads at times to 
violation of no-arbitrage relations for bespoke portfolios across strikes or maturities. 

Further problems with the use of base correlation mapping rules have to do with market 
incompleteness.
Market prices of standard tranches provide only information on {\it average} risk-neutral 
default dependencies for a well-diversified (across different industrial sectors) index 
portfolio, and {\it not} information about 
default correlations specific to different industrial sectors. This implies that the base 
correlation method within
a one-factor Gaussian copula framework does not capture sector concentration risk, and 
thus can
be potentially dangerous to use for bespoke portfolios whose sector or/and geographical 
composition is materially different from that of standard portfolios. 
 For a further discussion of practical difficulties arising with the use of base correlation 
 approach for pricing bespoke tranches, see  e.g. \cite{SocGen}, \cite{HW2006} and \cite{RS}.

\subsection{Relation to previous literature}

Our model combines elements of both bottom-up and top-down approaches. 
Here we would like to give a brief account of literature most relevant to our approach.

In a typical bottom-up model, we start at the view of a credit portfolio as a collection of 
separate exposures ("single names"), whose default dependencies are then introduced
via a factor copula in structural models, or via a multi-variate intensity process.
Any consistent multi-factor bottom-up model, with parameters tuned to roughly 
match the observed data, can be used as the ``prior'' model in our approach. Examples 
inlcude, in particular, the  
Gaussian copula model of Li \cite{Li}, a multi-factor version of the RFL model \cite{AS}, or 
reduced-form models, see e.g.
Duffie and Garleanu \cite{DG}, or Chapovsky et al \cite{Chapovsky}.

Once the prior model is chosen, we switch to a top-down paradigm, concentrating on the 
dichotomic representation of portfolio losses (see above), and abstracting from the initial
single-name picture. The idea of modeling credit portfolios in a top-down manner is originally 
due to Giesecke and Goldberg \cite{GG}, Sidenius, Peterbarg and Andersen \cite{SPA}, and 
Sch{\"o}nbucher \cite{Schon}. Our approach resembles the BSLP model of Arnsdorf 
and Halperin \cite{AH} in the 
sense that transition from the initial (``prior'') distribution to a 
``true'' (calibrated) one is 
achived via
a set of multiplicative loss-dependent factors applied directly to the portfolio loss 
distribution,
though a particular realisation of this idea in the present framework is different. 

Effectively, our procedure amounts to constructing loss distributions
for both the index and bespoke portfolio in a way {\it implied} by observed prices of standard tranches. Thus our method belongs in the class of implied loss distribution 
approaches which have become quite popular among practitioners in recent years.
Examples of such approach include e.g. the implied copula 
model of Hull and White \cite{HW2006}, \cite{HW2008}, or the 
factor model of Inglis and Lipton \cite{IL}. Unlike these authors, we 
work in a multi-factor setting that starts with a 
usual factor model, and consistently apply 
information-theoretic methods in order to construct implied loss distribution, while 
e.g. Hull and White adopt smoothness criteria as the main tool in their method.
 
 In employing a multi-factor framework and using 
 information-theoretic (entropy-based) methods for calibration, our approach 
is akin to that taken 
 by Rosen and Saunders \cite{RS}. They develop an entropy-based calibration method for a 
class of multi-factor bottom-up models, where adjustment to market prices amounts
 to calculation of implied  market factor distribution. 
  Unlike the latter authors, we use a top-down entropy calibration where
we calibrate to losses in tranches and suitably chosen sub-portfolios (see below), 
but not to individual names. This choice is made for the sake of simplicity and 
efficiency of implementation, and to facilitate an easy transition to a dynamic setting.  
On the implementation side, Rosen and Saunders employ a Monte Carlo scheme which is 
a preferred method when the number of market factors is three or more, while 
in our model we settle for a practically-oriented two-factor 
framework\footnote{As discussed by Rosen and Saunders \cite{RS}, a practical difficulty with 
using more market factors is that market prices 
  convey virtually no information on distributions of sector-specific factors.}, and 
design an efficient lattice-based 
scheme for calibration and pricing, which is comparable in 
performance to the BSLP model of Ref.\cite{AH}. 
The main difference of our present framework from more traditional top-down approaches
is that for pricing of bespoke tranches, we need a finer resolution of portfolio loss 
scanarios, which is achieved via calculation of joint loss distributions of sub-portfolios
of credit indices.
In concentrating on the dynamics of losses in sub-portfolios of index portfolios, our 
approach is similar in spirit to
a multi-portfolio top-down model of Zhou \cite{Zhou} (see also \cite{BSLPpartition}).  
In order to calculate 
loss distributions of sub-portfolios, Zhou relies on the random thinning 
technique similar to that of Giesecke and Goldberg \cite{GG}, and  Halperin and 
Tomecek \cite{HT}. However, his model does not 
employ a factor framework, which 
in our opinion is very useful, both conceptually and computationally,
for modeling bespoke portfolios that 
have to be mapped onto more than one index portfolio.  
The approach taken in the present paper (both the parametrization and 
calibration method) is different from those used in \cite{AH}, \cite{Zhou} and \cite{HT}.  

\section{Implied Multi-Factor Model at a glance }

In this section we provide a high level, qualitative overview of our approach.
All technical details are left for the next section. Here our task is to 
introduce different key components of our framework and explain how they help to 
address various deficiencies of more traditional approaches to credit portfolio modeling
in general, and the bespoke mapping problem in particular.


\subsection{No-arbitrage}

No-arbitrage conditions ensure that losses can only increase over time, and are  
clearly among the most important requirements
for a consistent bespoke model. The way no-arbitrage is enforced depends on 
the model. For example, in a dynamic model it can be imposed on the loss {\it process},
while for single-period models, it is usually formulated as conditions on 
{\it expectations} of future losses as functions of the time horizon and 
loss level. In 
particular, expected loss for a tranche
or a portfolio should increase with the time horizon, ensuring no-arbitrage across time.
No-arbitrage across loss levels is enforced as long as 
the expected loss for first loss (equity) tranche is an increasing and concave 
function of the strike. This ensures that the portfolio loss distribution 
is non-negative at any loss level. 
Recall here that the base correlation methodology, the 
current market standard, leads to occasional violations of no-arbitrage in both 
the strike and time 
dimension.  Such failure to ensure no-arbitrage can be traced back to the fact that 
in the base correlation  approach, bespoke tranches are priced by 
interpolation/extrapolation
in the "wrong" space (the correlation space), where conditions of no-arbitrage are hard to 
check or enforce. On the contrary, in our model no-arbitrage across strike is ensured by 
the fact that the bespoke price is calculated directly in the loss space using  
probabilistic arguments without any need of interpolation in an auxiliary correlation space,
with the loss density being non-negative by construction as long as our ``prior'' model
$ Q $ is arbitrage-free.  
This follows simply from the fact that in our model the ``true'' measure $ P $ is 
found by minimization of a information-theoretic ``distance'' (KL-divergence, see below in 
Sect.3) between $ P $ and $ Q $, which is finite
as long as two measures $ P $ and $ Q $ are equivalent. Therefore, as long as 
the prior distribution $ Q $ is arbitrage-free, the "true" distribution $ P $ obtained using 
minimization of the KL-divergence is guaranteed to be arbitrage-free as well.



The question of arbitrage across time is somewhat more subtle. Our model comes in 
two versions: one-period and multi-period.
In the former, no-arbitrage is guaranteed by construction across strikes, but not 
necessarily across time (though the latter is found to hold {\it a posteriori} in our 
numerical experiments). The dynamic version is free of arbitrage across both strikes and 
time by construction. No-arbitrage 
across time is ensured as long as we pick an arbitrage-free prior model for the portfolio 
loss process, see below.

\subsection{Multi-factor structure}

Most of the models currently in use in the industry are one-factor models, where default 
dependence between names in a portfolio arises due to dependence of 
individual defaults on a one-component 
"market" factor $ Z $ that is common for all names in the portfolio.
Such approach, expressing default dependencies as a 
result of dependence on a single common factor,  may be reasonable for a well-diversified 
portfolio such as CDX.IG.NA. 

\begin{figure}
\begin{center}
\includegraphics[width=13cm, height=90mm]{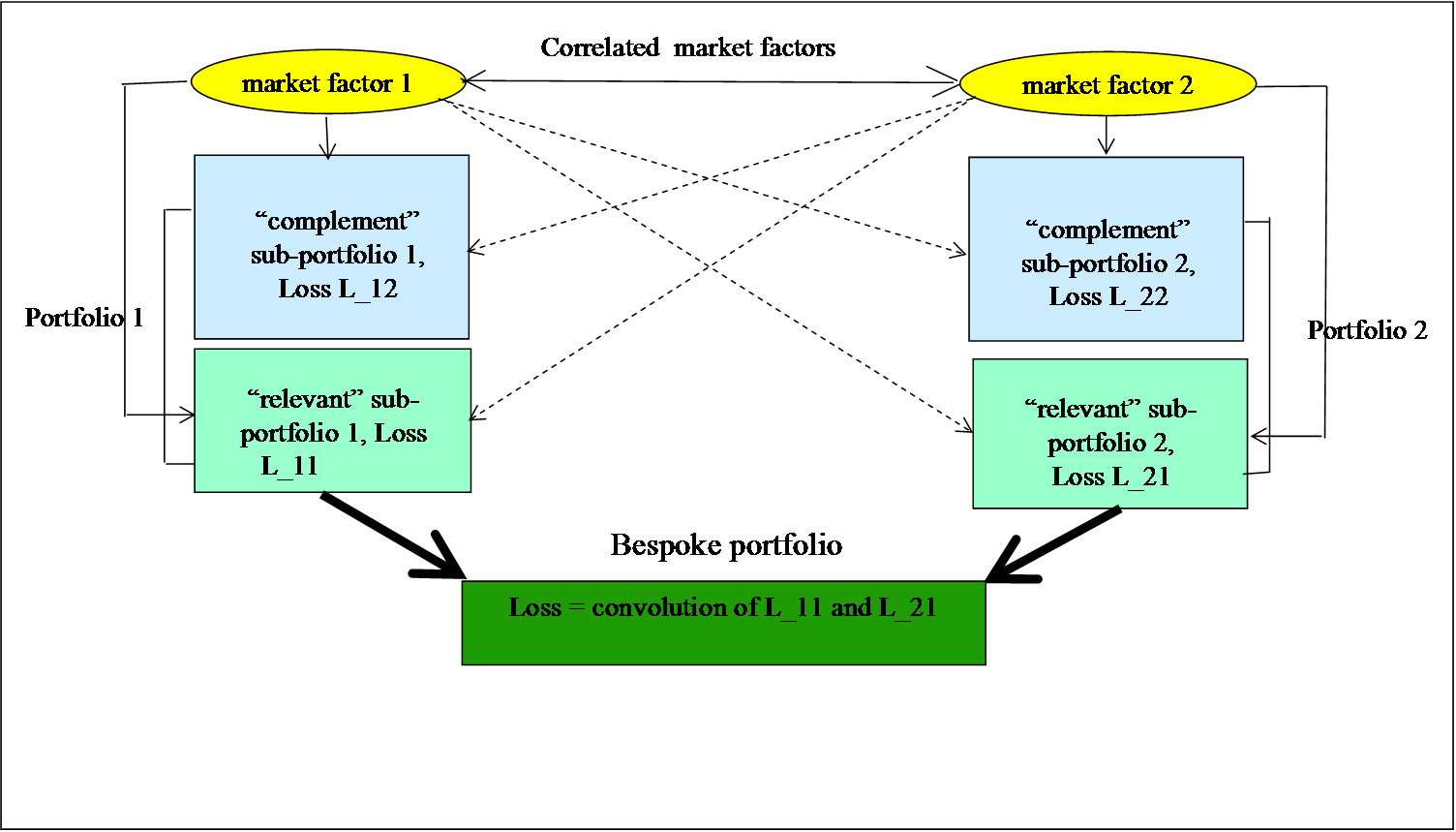}
\caption{Layout of the Implied Multi-Factor Model (IMFM).} 
\label{fig:IMFM_layout}
\end{center}
\end{figure}

However, it is by no means clear that the same one-factor framework 
can be applied to bespokes whose composition is   
materially different from that of an index portfolio, like our $ I2$-type bespokes 
introduced above. It appears that for 
such bespokes, a more realistic framework
that takes into account diversification effects (i.e. qualitative differences 
in portfolio composition), should include several  market factors. As a minimum 
requirement, we should 
have two (dependent) factors $ Z_1 $ and $ Z_2 $, that could be thought of as  
a "US factor" and "Europe factor", or a "geographic factor" and "industry 
factor", see Fig.~\ref{fig:IMFM_layout}.
A two-factor framework thus seems to provide a minimal complexity level for a 
bespoke model needed for all but simplest bespoke portfolios that only need to be mapped onto
one credit index. Given that market-implied prices of standard index tranches contain 
virtually no
information on factors corresponding to separate sectors beyound a single factor 
corresponding to a 
diversified credit index as a whole\footnote{See Rosen and Saunders \cite{RS} 
for a discussion on 
this point.},
we believe that a two-factor specification may simultaneously serve as the 
upper ``complexity bound''
for a bespoke model to be useful and manageable in practice\footnote{More factors 
might be needed for more complex bespokes referencing more than two credit indices,
e.g. for $ I3 $-type bespokes we would need three factors, etc.}.
We therefore stick to a 
two-factor
formulation in what follows, and leave generalization for Sect.5.  


\subsection{Types of bespokes that can be priced with IMFM}

As mentioned above, we want to be able to price arbitrary bespoke tranches which, 
by nature of their composition, should be mapped onto more than one index portfolio.
We achieve this by constructing a certain hierarchy of "bespoke complexity". 
We first address a simplest bespoke portfolio: a bespoke that is made of 
"chunks" of index portfolios. Assume we have two index portfolios $ \Pi_1 $ and 
$ \Pi_2 $  corresponding e.g. to the CDX.NA.IG and iTraxx Europe Main credit portfolios.
Our simplest bespoke portfolio is made of names making up certain sub-portfolios
(hereafter ``relevant'' sub-portfolios)  
$  \Pi_{11} $ 
and $ \Pi_{21} $ of index portfolios, respectively, while all names 
from complement 
sets  $ \Pi_{12} = \Pi_1 \backslash \Pi_{11} $ and 
$ \Pi_{22} = \Pi_2 \backslash \Pi_{21} $ will be collectively referred to 
as "complement" sub-portfolios (see Fig.~\ref{fig:IMFM_layout}). We 
will refer to such bespokes made of names from two, three
etc. indices as $ I2, I3, \ldots $ portfolios. 
One example of a $ I2 $ portfolio would be a bespoke portfolio with 100 names 
with 50 US names from the CDX.NA.IG index portfolio, and 50 European names 
from iTraxx Europe index portfolio.  

Analysis in the following Sects. 3 and 4 will concentrate on the pricing of $ I2 $-type
bespoke portfolios. 
This setting forms our basic case which will be worked out in great details. 
Treatment of more complex bespoke portfolios (e.g. portfolios made of chunks of three and 
more indices, or 
portfolios containing names not belonging to any index) will be presented as suitable 
generalizations of our basic setting in Sect.5.

\subsection{Other applications: credit exotics and CVA on bespoke tranches}

In addition to the ability to price bespoke CDO tranches on portfolios mapped onto
more than one credit index, our model has, for its dynamic
version, other interesting applications as well. First, more exotic portfolio 
derivatives, such as e.g. tranche options or forward-starting tranches, can be 
priced using a loss
lattice with a backward recursion method, similar to how it is done in the BSLP model
\cite{AH}. Second, the dynamic version of our multi-factor framework can be used
for counterparty risk management of CDO tranches, in particular, for calculation of 
credit valuation adjustments (CVAs) on bespoke tranches. Again, a lattice formulation
enables a quick and efficient implementation of the computational algorithm for 
this case.

\subsection{Computational efficiency: a top-down approach}

A multi-factor credit portfolio model of a traditional buttom-up type can run into 
numerical challenges rather quickly. Indeed, in consistent models for tranche 
prices such 
as RFL (which is a single-factor model), calibration takes around 5-30 min,
depending on the model, efficiency of implementation and the particular 
calibration dataset. When 
moving to a two-factor framework with a double number of names, calibration time is 
expected to grow by at least an order of 
magnitude\footnote{Let us assume for simplicity that the number of steps of an 
optimization 
algorithm (e.g. a gradient-based one) needed to calibrate the model is proportional
to $ N_{tranche} $ - the number of tranches in the calibration set. Then the total 
time needed to calibrate the model scales as $ T \sim N_{names}^2 N_{market}^{D}
N_{tranche} $, where the first factor is due to $ \sim N_{names}^2 $ operations needed 
to perform convolutions to calculate conditional loss 
distributions, while the second factor
is due to fact that this calculation should be repeated $ N_{market}^{D} $ time, where
$ N_{market} $ is the number of discretization points for one component of the factor, 
and $ D $ is the number of component. 
If we use e.g. a 10-point Gaussian quadrature for integration over the market factor
(i.e. $ N_{market} = 10 $), when we proceed from one factor setting with 
$ N_{names} = 125 $ (i.e., one index), $ N_{tranche} = 6 $, $ D = 1 $ to 
a two-factor setting and two index portfolios, the 
computational time increases by a factor $ 4 N_{market} = 40 $.}
, which might become computationally infeasible on a single PC.

We choose to address these potential issues by adopting a version of the top-down approach
to credit portfolios  modeling. To introduce our setting,   
let us return to the case of a simplest $ I2 $-portfolio of the previous section.
Recall that this is a bespoke composed of two "chunks" of two index portfolios (out of 
the total of four "chunks"). 
 Let $ X_{11} $ and $ X_{21} $ be stochastic
losses for "relevant" sub-portfolios  $  \Pi_{11} $ 
and $ \Pi_{21} $, respectively, with $ X_{12} $ and $ X_{22} $ 
being losses in complement sub-portfolios $ \Pi_{12} = \Pi_1 \backslash \Pi_{11} $ and 
$ \Pi_{22} = \Pi_2 \backslash \Pi_{21} $. Then we have the 
following obvious relations for the 
total losses $ X_1 $ and $ X_2 $ for portfolios  $\Pi_1 $ 
and $ \Pi_2 $:
\beq
X_1 = X_{11} + X_{12} \; \; , \; \;  X_2 = X_{21} + X_{22}
\eeq 
The top-down specification of our approach amounts to the fact we only keep track of 
cumulative sub-portfolio losses $ \vec{X}_1 = \left(X_{11}, X_{12} \right) $ 
and 
$ \vec{X}_2 = \left(X_{21}, X_{22} \right) $ (we use hereafter a compact vector 
notation) , but {\it not} default states of individual names
in either the "relative" or "complement" sub-portfolios. 
As long as the composition of our bespoke 
is as described above\footnote{Generalizations will be introduced below.},  
the joint 
distribution of sub-portfolio
losses $ X_{11} $ and $X_{21} $ is all we need to price the bespoke.    

The latter is easy to demonstrate. Consider a given time horizon $ T $, and 
assume for the moment that we have somehow managed 
to calibrate this market factor distribution together with conditional
joint loss distributions of sub-portfolios of the index portfolios  
$ P_1 \left(X_{11}, X_{12} | \vec{Z} \right) $ 
and $ P_2 \left(X_{21},X_{22} | \vec{Z} \right) $ for this horizon.
We can now easily price any tranche 
referencing the bespoke portfolio $ \Pi_B = \Pi_{11} \cup \Pi_{21} $. Indeed, as 
losses in two portfolios are conditionally independent, the 
conditional loss distribution is easily
found by convolution
\beq
\label{condLossDistrBespoke}
P^B ( X_{11} + X_{21} = x | \vec{Z}) = \int d y \,  P_1 \left(y | 
\vec{Z} \right) 
 P_2 \left(x - y | \vec{Z} \right)
\eeq
where conditional loss distributions  
$  P_1 \left(X_{11} | \vec{Z} \right)  $ and 
$  P_2 \left(X_{21} | \vec{Z} \right) $ are obtained by marginalization of 
$ P_1 \left(X_{11}, X_{12} | \vec{Z} \right) $ and
 $  P_2 \left(X_{21},X_{22} | \vec{Z} \right)  $ over $ X_{12} $ and $ X_{22} $,
respectively. Assuming that the common market factor is a continuous random variable
with pdf $ P_{Z}(\vec{z}) $, 
the unconditional loss distribution is then obtained as follows:
\beq
\label{uncondLossDistrBespoke}
p^B (x,T) = \int P^ B (x | \vec{z}) P_Z(\vec{z}) d \vec{z}
\eeq
Once the loss distribution for the bespoke portfolio is calculated on a grid $ \{ T_i  \} $ 
of time horizons, any bespoke tranche is priced by standard formulae using numerical 
integration, see below in Sect.3.3.

\section{IMFM in a single-period setting}

In this section we present the Implied Multi-Factor  Bespoke Model (IMFM) in 
a simplified single-period setting, where we only deal with the losses of index 
portfolios and their sub-portfolios at a single time horizon $ T $. Once 
worked out, this will serve 
as a seed for a dynamic multi-period extension which is introduced in the next section.


\subsection{A two-factor CreditMetrics prior}


As was mentioned above, while our formulation is straightforward to generalize to an arbitrary
number of market factors, our basic formulation will deal with a specific case of two market 
factors.
In component notation, the latent variable for the $i$-th obligor in the $ k$-th index 
portfolio takes the following form
\beq
\label{twoComponent}
A_i^{(k)} = \beta_{i1}^{(k)} Z_1 + \beta_{i2}^{(k)} Z_2 + \sqrt{1 - 
\left( \beta_{i1}^{(k)} \right)^2 - \left( \beta_{i2}^{(k)} \right)^2 
- 2  \rho \beta_{i1}^{(k)} \beta_{i2}^{(k)}} \varepsilon_i
\eeq
Here $ Z = (Z_1, Z_2) $ is a two-dimensional Gaussian random variable with 
means 0, variances of one, and correlation $ \rho $. 

While parameters of such model can be estimated from time series of spreads or 
defaults, we leave this task for future work, and instead 
adopt in this paper a simple parametrization that ensures consistency with a one-factor 
Gaussian copula framework. Assuming that the latter model is already estimated, our 
parametrization guarantees that no further work for parameter estimation is required. 

We concentrate on a basic setting of two index portfolios, $ \Pi_1$ and $ \Pi_2 $. We 
assume that
\beq
\label{simplePar}
\beta_{i2}^{(1)} = \alpha \beta_{i1}^{(1)} \; \; , \; \; 
\beta_{i1}^{(2)} = \alpha \beta_{i2}^{(1)}
\eeq
In words, we assume that for each name, its factor loading to the "foreign" factor is 
a fixed proportion of its factor loading to its "domestic" factor. It is easy to 
see that, 
as long as we view portfolio $ \Pi_ 1 $  ( $ \Pi_2 $) in isolation from portfolio $ \Pi_2 $ 
(resp. $ \Pi_1$), the resulting model is identical to a one-factor model with factor 
loadings $ b_i^{(1)} $ (or $ b_{i}^{(2)} $) provided we set 
\bea
\label{linkOneFactor}
\beta_{i1}^{(1)} &=& \frac{b_i^{(1)}}{\sqrt{1 + 2 \alpha \rho + \alpha^2}} \nonumber \\
\beta_{i2}^{(2)} &=& \frac{b_i^{(2)}}{\sqrt{1 + 2 \alpha \rho + \alpha^2}}
\eea
Note that this framework slightly generalizes a more conventional hierarchical structure
which is recovered with the present formalism in the limit $ \alpha = 0 $.

Our parametrization (\ref{simplePar}) ensures that pairwise asset correlations for 
names in the same portfolio are the same as in the one-factor model irrespective
of values of $ \rho $ and $ \alpha $, e.g. for portfolio $ \Pi_1 $ we obtain
\beq
\label{pairwiseTheSame}
\rho_{ij}^{(11)} = \beta_{i1}^{(1)} \beta_{j1}^{(1)} ( 1 + 2 \alpha \rho + \alpha^2) 
= b_{i}^{(1)} b_{j}^{(1)}      
\eeq
On the other hand, for two names in different portfolios we find
\beq
\label{pairwise2}
\rho_{ij}^{(12)} = \beta_{i1}^{(1)} \beta_{j2}^{(2)} ( \rho(1 + \alpha^2) + 2 \alpha ) 
= b_{i}^{(1)} b_{j}^{(2)}  \frac{ ( 1 + \alpha^2) \rho + 2 \alpha }{ 1 + \alpha^2 + 2 \alpha
\rho} \geq \rho b_{i}^{(1)} b_{j}^{(2)} 
\eeq
Note that equality in the last expression follows in the limit $ \alpha = 0 $ 
corresponding to a hierarchical model. 
Therefore, for a fixed $ \rho $, a non-zero value $ \alpha > 0 $ enhances inter-sector 
correlations\footnote{Note that while any given level of pairwise correlations obtained 
with a non-zero value of $ \alpha $ can be reproduced with $ \alpha = 0 $ and a higher value
of $ \rho $, higher-order correlations (and hence tranche prices) in these two cases are 
different.}.

In the implementation of the model, we discretize the market factor as follows.  
Using vector notation
$ \vec{Z} = \left(Z_1, Z_2 \right) $, the range of possible values of $ \vec{Z} $ is 
discretized to a 2D grid 
\beq
\label{grid}
 \vec{Z}_{\vec{m}} = \left( z_{m_1} , z_{m_2} \right)
\eeq 
where  $ m_1 = 1, 
\ldots, N_{m1} $ and $ m_2 = 1, \ldots, N_{m2} $, and 
\beq
\label{vecN}
\vec{m} = \left( m_1, m_2 \right) \;, \; \;  m_1, m_2 \in Z^{+} 
\eeq
is a 2D vector of integers. Corresponding discretized probabilities of 
realizations of these
values of the market factor on the grid $ \left\{ z_{m_1} , z_{m_2} \right\} $ read 
\beq
\label{hn}
h_{\vec{m}} \equiv h_{m_1, m_2} = P \left[ Z_1 = z_{m_1}, Z_2 = 
z_{m_2} \right] = P \left[ \vec{Z} = \vec{Z}_{\vec{m}} \right]
\eeq 
The difference the prior and ``true'' discretized market factor distributions 
thus amounts to different choice for weights $ \{ h_{\vec{m}} \} $.
In the numerical examples to follows, we  
use a low number (between 10 and 20) of discretization points for 
each factor.

 
\subsection{IMFM calibration via minimization of KL-divergence}

Our problem is to calculate the joint distribution of the 6 random variables
$ \left(\vec{Z}, \vec{X}_1 , \vec{X}_2 \right) $ which is {\it implied} by the observed
set of tranche price quotes. More exactly, we calibrate our model to 
a set of tranche expected losses at different maturities, which are obtained from 
market tranche prices using an arbitrage-free interpolation method. For the latter, 
one option is to use a spline-based method where no-arbitrage conditions are enforced 
as constraints, or using a top-down model such as BSLP \cite{AH} that does such 
interpolation internally. We do not expand on this procedure here, and 
instead refer the reader to the literature.
 
 The implied joint distribution of $ \left(\vec{Z}, \vec{X}_1 , \vec{X}_2 \right) $
 is calculated as a minimal distortion (in the sense of information entropy, see below) 
 of a "prior" distribution (see below for possible choices for the prior), with the 
distortion just sufficient in order to match a given set of  
tranche expected losses $ EL_{ki} $ , where $ i = 1,2 $ stands for the 
credit index portfolio, and $ k = 1, \ldots , N_i $ 
 enumerates tranches referencing the $ i$-th portfolio. In 
information theory (see e.g. \cite{CT}), 
 the "distance"-type measure for two continuous distributions of 
 a random variable $ Y $ with densities $ P_{1}(y) $ and $ P_{2}(y) $  is given by the celebrated Kullback-Leibler relative 
 entropy (KL-divergence)
 \beq
 \label{KL}
 D \left[ P_2 || P_1 \right] = \int dy P_2(y) \log \frac{P_2(y)}{P_1(y)} 
 \eeq
 It can be easily checked using Jensen's inequality that $ D[ P_2 || P_1 ] $ is 
non-negative for all pairs $ \left(P_1, P_2 \right) $, and reaches zero only 
 when $ P_2 = P_1 $. Note that for discrete distributions, integration in (\ref{KL})
  should be substituted by summation. Having this in mind, in what follows we will
  keep for a while the continuous notation even for discrete distributions.
 
 In our setting, variable $ y $ appearing in (\ref{KL}) is 
6-dimensional (6D)\footnote{Higher dimensionality is obtained with either or both of finer portfolio partitioning or increased
 dimensionality of the market factor. The case $ D=6$ seems to be the case of minimal 
 dimensionality $ D $ needed for the bespoke problem.}:
 \beq
 \label{6dim} 
 Y = \left(\vec{Z}, \vec{X}_1 , \vec{X}_2 \right) \equiv \left( \vec{Z}, \vec{X} \right)
 \eeq
 where $ \vec{X} = \left( \vec{X}_1, \vec{X}_2 \right) $ is a 4D vector storing all sub-portfolio losses in both index portfolios.
 The generalized KL-divergence
 for multi-dimensional distributions takes the same form as (\ref{KL}), provided integration
 (or summation) is extended over the whole  $D$-dimensional space. 
 
In what follows we assume that we pick a particular
"prior" model $ Q $ for the joint risk-neutral distribution of 
$ \left(\vec{Z}, \vec{X}_1 , \vec{X}_2 \right) $:
 \beq
 \label{Q}
 Q = Q \left(\vec{Z}, \vec{X}_1 , \vec{X}_2 \right) = Q \left( \vec{Z}, \vec{X} \right)
 \eeq
We note that a few choices for the latter are popular
 among practitioners. One option is to use a uniform prior distribution 
corresponding to the state of maximum uncertainty about the value of  $ Y $. 
Another, and often more attractive choice is to use a prior measure
corresponding to a specific parametric model.  
We will adopt a 2D Gaussian copula model with factor loadings defined as in 
(\ref{simplePar}) and (\ref{linkOneFactor}) 
as the prior model in our approach\footnote{Alternatively, we could use the same 
model calibrated to historical data.}.  
 
Following the Minimum Cross-Entropy Method (MCE)\footnote{The MCE method generalizes the 
famous Maximum Entropy (MaxEnt) method. The latter is recovered 
from the former when one adopts a uniform prior. While the MaxEnt and MCE methods
have been widely used since early 1980s 
in such fields as image processes, speech recognition etc.
see e.g. \cite{CT}, its first uses in the context of option modeling
appeared around 1995-1996 in a series of papers by Avellaneda et al \cite{Avella}, 
Buchen and Kelly \cite{BK},
Gulko \cite{Gulko}, and Stutzer \cite{Stutzer}. For more recent applications for 
modeling credit portfolios, see e.g. Halperin \cite{HalperinCRS2006}, 
Vacca \cite{Vacca} and  Veremeyev et al 
\cite{Uryasev}. }, we 
seek an implied joint risk-neutral distribution $ P\left(\vec{Z}, \vec{X} \right) $ 
that minimizes the 
 KL-divergence between the ``true'' distribution $  P\left(\vec{Z}, \vec{X} \right)  $ and 
the prior distribution $  Q\left(\vec{Z}, \vec{X} \right)  $:
 \beq
 \label{KL1}
 D \left[ P\left(\vec{Z}, \vec{X} \right) || Q\left(\vec{Z}, \vec{X} \right) \right] = \int d \vec{Z}  d \vec{X} 
 P \left(\vec{Z}, \vec{X} \right) \log \frac{
 P\left(\vec{Z}, \vec{X}  \right)}{Q\left(\vec{Z}, \vec{X} \right)}
 \eeq
 (here $ d \vec{X} = \prod_{i=1,2} d \vec{X}_i $, and we use the integral notation for 
 summation over the possible values of $ \vec{Z} $), subject to pricing constraints
 \beq
 \label{constraints}
 \int d \vec{Z}  d \vec{X}   P \left(\vec{Z}, \vec{X} \right)
 F_{ik} \left(\vec{X}  \right) = EL_{ik}\; , \; \; i = 1,2\; , \; k = 1\ldots, N_i, N_i+1,
 N_i + 2
 \eeq
 where for a given $ i $, the first $ k = 1, \ldots, N_i $ components of 
 the generalized payoff functions $ F_{ik} \left(\vec{X} \right) $ stand for the payoffs
 of the $ k$-th tranche referencing the $ i$-th index portfolio, while the last $ (N_i+1) $-th and 
 $ (N_i + 2)$-th components correspond to the payoffs (total expected losses) for sub-portfolios $ \Pi_{i1} $ and  $ \Pi_{i2} $, respectively:
\bea
\label{payoffs}
F_{ik}(\vec{X}) = 
\left\{ \begin{array}{lll}
(X_{i1} + X_{i2} - K_{i,k-1})^{+} - (X_{i1} + X_{i2} - K_{ik})^{+} & \mbox{$k = 1, \ldots, N_i$} \nonumber \\
X_{i1} & \mbox{$k = N_i + 1$}  \\  
X_{i2} & \mbox{$k = N_i + 2$} 
\end{array}
\right.
\eea
where $ \{ K_{ik} \} $ is a set of standard strikes for the $i$-th index portfolio. 
Generalized expected losses $ EL_{ik} $ are defined using a similar convention.
Eqs.(\ref{constraints}) plus the normalization condition $ \int d \vec{Z}  
d \vec{X}   P \left(\vec{Z}, \vec{X} \right) = 1 $ 
constitute the full set of constraints imposed in our model.   
   
We follow the factor framework, i.e. we assume that $ \vec{Z} $ is 
an unobservable "market" factor such that conditional on the value of $ \vec{Z} = 
\vec{Z}_{\vec{m}} $,
all individual defaults in both index portfolios $ \Pi_1 \, , \Pi_2 $ become independent.
Correspondingly, the joint probability of $ \left(\vec{Z}_{\vec{m}}, \vec{X} \right) $ can
be written as a product of the "market factor" probabilities $ h_{\vec{m} } $ (see 
Eq.(\ref{hn})) and conditional loss distributions $ P_i \left( \vec{X}_i | \vec{m} \right) 
\equiv P_i \left( \vec{X}_i | \vec{Z} = \vec{Z}_{\vec{m}} \right) $ ($ i = 1,2 $)
of losses in two index portfolios:
\beq
\label{jointFactor}
P \left(\vec{Z} = \vec{Z}_{\vec{m}}, \vec{X}\right) = 
h_{\vec{m}} P \left( \vec{X} | \vec{m} \right) = 
h_{\vec{m}} \prod_{i=1,2} P_i \left(\vec{X}_i | \vec{m} \right)
\eeq
Similarly, we write the prior model in the factor form:
\beq
\label{jointFactorPrior}
Q \left(\vec{Z} = \vec{Z}_{\vec{m}}, \vec{X} \right) = 
g_{\vec{m}} Q \left( \vec{X} | \vec{m} \right) = 
g_{\vec{m}} \prod_{i=1,2} Q_i \left(\vec{X}_i | \vec{m} \right)
\eeq
Substituting (\ref{jointFactor}), (\ref{jointFactorPrior}) into (\ref{KL1}), the KL-divergence can be written as 
follows\footnote{Here we switch to discrete summation over $ \vec{Z} $}:
\beq
\label{KL2}
 D \left[ P\left(\vec{Z}, \vec{X} \right) || Q\left(\vec{Z}, \vec{X} \right) \right] 
 = \sum_{\vec{m}} h_{\vec{m}} \log \frac{ h_{\vec{m}}}{g_{\vec{m}}} 
 + 
 \sum_{i=1,2} \sum_{\vec{m}} h_{\vec{m}}  \int d \vec{X}_i  
P_i \left(\vec{X}_i | \vec{m} \right) \log \frac{
P_i\left(\vec{X}_i | \vec{m} \right)}{Q_i\left(\vec{X}_i | \vec{m} 
\right)} 
\eeq
The problem of minimization of the KL-divergence of distributions $ P
\left(\vec{Z}, \vec{X} \right) $ and $ Q \left(\vec{Z}, \vec{X} \right) $ subject to constraints 
(\ref{constraints}) can be solved in one of two ways. The first one is 
a direct functional minimization of KL-divergence in form (\ref{KL1}) viewed as a functional of the joint probability $ P \left( \vec{m}, \vec{X} \right) = 
P \left(\vec{Z}_{\vec{m}}, \vec{X} \right) $. The second, and equivalent, way is to use the expression (\ref{KL2}) and minimize it jointly by viewing it as a functional of 
{\it a priori}\footnote{i.e. prior to taking into consideration constraints 
(\ref{constraints}).} independent distributions
$ h_{\vec{m}} $ and $ P_i \left( \vec{X} | \vec{m} \right) $. 

It is instructive to start with the second approach.
We pose the variational optimization problem with the following Lagrangian functional:
\bea
\label{Lagrange}
\mathcal{L}' & = &  \sum_{\vec{m}} h_{\vec{m}} \log \frac{h_{\vec{m}}}{g_{\vec{m}}}  +  
\sum_{i, \vec{m}} h_{\vec{m}}  \int d \vec{X}_{i} \, P_i
\left(\vec{X}_i | \vec{m} \right)
\log \frac{ P_i\left(\vec{X}_i | \vec{m} \right) }{ Q_i\left(\vec{X}_i | \vec{m} \right)} \nonumber \\ 
&+&  \sum_{i, k} \frac{1}{2 \sigma_{ik}^2}  
\left( \sum_{\vec{m}} h_{\vec{m}} \int d \vec{X}_i \,
P_i\left(\vec{X}_i | \vec{m} \right)   F_{ik}(\vec{X}) - EL_{ik} \right)^2 
\eea
which should be minimized with respect to $ h_{\vec{m}} $, $ 
P_1 \left( \vec{X}_1 | \vec{m} \right) $ and $ P_2 \left( \vec{X}_2 | \vec{m} \right) $.
Here the first two terms enforce minimization of the KL-divergence 
between the ``true'' and prior joint distributions of the market factor and losses, 
while the last term enforces matching the tranche pricing data in the least square 
sense\footnote{For brevity, we have omitted additional
terms in (\ref{Lagrange}) corresponding to normalization constraints for  
distributions of interest, however they will always be kept in mind in the 
calculation to follow.}. 

Note the relative importance of matching the data versus minimization of the KL ``distance'' to 
the prior distribution is controlled by parameters $ \sigma_i $. In particular, when these 
parameters are large, the ``true'' distribution is very close to the prior, while the quality of fit 
is poor. In the opposite limit $ \sigma_i \rightarrow 0 $, one imposes an 
exact matching of pricing constraints. In this limit, the suitable Lagrangian reads,
instead of (\ref{Lagrange}),
\bea
\label{LagrangeExactConstraints}
\mathcal{L}' & = &  \sum_{\vec{m}} h_{\vec{m}} \log \frac{h_{\vec{m}}}{g_{\vec{m}}}  +  
\sum_{i, \vec{m}} h_{\vec{m}}  \int d \vec{X}_{i} \, P_i
\left(\vec{X}_i | \vec{m} \right)
\log \frac{ P_i\left(\vec{X}_i | \vec{m} \right) }{ Q_i\left(\vec{X}_i | \vec{m} \right)} \nonumber \\ 
&+&  \sum_{i, k} \lambda_{ik}  
\left( \sum_{\vec{m}} h_{\vec{m}} \int d \vec{X}_i \,
P_i\left(\vec{X}_i | \vec{m} \right)   F_{ik}(\vec{X}) - EL_{ik} \right) 
\eea 
where $ \lambda_{ik} $ are Lagrange multipliers. 
While such formulation corresponds to 
a more conventional version of the MCE/MaxEnt methods, it may potentially lead to unstable 
calibration and 
overfitting, or even to no solution at all if constraints are incompatible. We therefore prefer the 
``soft constraint'' version (\ref{Lagrange}), where 
a compromise between robustness and quality of fit is expected to be achieved for some 
intermediate values of $ \sigma_i $.

Unlike the ``classical'' MCE setting (\ref{LagrangeExactConstraints}),
a straightforward functional optimization of Lagrangian 
(\ref{Lagrange}) is diffucult because 
of its non-linearity. Fortunately, a simple method is available that allows one to get 
rid of these non-linearities at the price of introducing auxiliary variables.
Following Ref.\cite{IH2002}, we consider the following 
modified Lagrangian
\bea
\label{modL}
\mathcal{L} & = &  \sum_{\vec{m}} h_{\vec{m}} \log \frac{h_{\vec{m}}}{g_{\vec{m}}}  +  
\sum_{i, \vec{m}} h_{\vec{m}}  \int d \vec{X}_{i} \, P_i
\left(\vec{X}_i | \vec{m} \right)
\log \frac{ P_i\left(\vec{X}_i | \vec{m} \right) }{ Q_i\left(\vec{X}_i | 
\vec{m} \right)} \nonumber \\ 
&+&  \sum_{i, k} \frac{1}{2 \sigma_{ik}^2} U_{ik}^2 + 
\sum_{i,k} \lambda_{ik}   
\left( U_{ik} - \sum_{\vec{m}} h_{\vec{m}} \int d \vec{X}_i \,
P_i\left(\vec{X}_i | \vec{m} \right)   F_{ik}(\vec{X}) + EL_{ik} \right) 
\eea 
It is easy to see that this modified Lagrangian is equivalent to the original one 
(\ref{Lagrange}) if, in addition to minimization with respect to distributions
$ h_{\vec{m}} $,  and $ 
P_1 \left( \vec{X}_i | \vec{m} \right) $ ( $ i = 1,2 $), we also vary it with respect 
to parameters  $ U_{ik} $ and $ \lambda_{ik} $. Indeed, if we first optimize 
(\ref{LagrangeExactConstraints}) with respect to $ \lambda_{ik} $, we obtain
\beq
\label{lambdaVariation}
U_{ik} = \sum_{\vec{m}} h_{\vec{m}} \int d \vec{X}_i \,
P_i\left(\vec{X}_i | \vec{m} \right)   F_{ik}(\vec{X}) - EL_{ik} 
\eeq
and back substitution in (\ref{modL}) leads to (\ref{Lagrange}). The same result is obtained if we
optimize wrt $ U_{ik} $ first, and $ \lambda_{ik} $ next.

Having established equivalence of (\ref{modL}) and (\ref{Lagrange}), we now want to proceed
differently with (\ref{modL}).
We start with minimization with respect to $ P_1 \left(\vec{X}_1 | \vec{m} \right)  $ and 
$ P_2 \left(\vec{X}_2 | \vec{m} \right)  $. We find 
\bea
\label{q1}
P_i\left(\vec{X}_i | \vec{m} \right)  &=& \frac{1}{Z_i (\vec{m}, \lambda)}
Q_i\left(\vec{X}_i | \vec{m} \right) 
e^{ \sum_{k} \lambda_{ik} \left( F_{ik} (\vec{X}_i)
- EL_{ik} \right) } \; , \; \; i = 1,2 \nonumber \\
Z_i( \vec{m}, \lambda) &=& \int d \vec{X}_i \, Q_i\left(\vec{X}_i | \vec{m} \right)  e^{ 
 \sum_{k} \lambda_{ik} \left( F_{ik} (\vec{X}_i)
- EL_{ik} \right) }
\eea
Interestingly, Eqs.(\ref{q1}) indicate that even though we start with the
factorized prior distributions $  Q_i \left( \vec{X}_i | \vec{m} \right) = 
Q_{i1} \left(X_{i1} | \vec{m} \right) Q_{i2} \left(X_{i2} | \vec{m} \right) $, we 
end up with non-factorizable conditional distributions  
$ P_i\left(\vec{X}_i | \vec{m} \right)  $ due to the fact that payoff 
functions $ F_{ik} $ depend on the sums $ X_{i1} + X_{i2} $  {\it non-linearly}. 
As discussed in  details in Appendix, this both makes sense intuitively and can 
be explained within Information Theory.
Therefore, while for the prior model losses in all four sub-portfolios $ \Pi_{11},
\Pi_{12}, \Pi_{21} $ and $ \Pi_{22} $ were independent conditional on the market factor, 
as a result of calibration, they become dependent within the same parent index portfolio,
while retaining independence across different index portfolios.

Let us now proceed with the solution of optimization
problem (\ref{Lagrange}).    
Substituting (\ref{q1})
back in the Lagrangian (\ref{modL}), we obtain
\beq
\label{Lagr2}
\mathcal{L} =  \sum_{\vec{m}} h_{\vec{m}} \log \frac{h_{\vec{m}}}{g_{\vec{m}}} - 
\sum_{\vec{m}} h_{\vec{m}} \sum_{i=1,2} \log Z_{i}(\vec{m}, \lambda) 
+ \sum_{i,k} \left( \frac{1}{2 \sigma_{ik}^2} U_{ik}^2 + \lambda_{ik} U_{ik} \right) 
\eeq
Minimizing this with respect to $ h_{\vec{m}} $, we find
\bea
\label{Hmn}
h_{\vec{m}} &=& \frac{1}{Z(\lambda)} g_{\vec{m}} 
\prod_{i=1,2} Z_{i} (\vec{m}, \lambda) \nonumber \\
Z(\lambda) &=& \sum_{\vec{m}}  g_{\vec{m}} 
\prod_{i=1,2} Z_{i} (\vec{m}, \lambda)
\eea
Minimization of (\ref{Lagr2}) with respect to $ U_{ik}$ gives
\beq
\label{Uik}
U_{ik} = - \lambda_{ik} \sigma_{ik}^2
\eeq
We next substitute the extremal values (\ref{Hmn}) and (\ref{Uik}) 
into (\ref{Lagr2}) to obtain
\beq
\label{Lagr3}
\mathcal{L} = - \log Z(\lambda) - \frac{1}{2} \sum_{ik} \lambda_{ik}^2 \sigma_{ik}^2 
\eeq  
The values of Lagrange multipliers should now be found numerically by maximization 
of (\ref{Lagr3}), or equivalently, minimization of the Lagrangian
\beq
\label{mini}
\mathcal{L'}(\lambda) =  \log Z(\lambda) + \frac{1}{2} \sum_{ik} \lambda_{ik}^2 \sigma_{ik}^2 
\eeq
This is a convex optimization problem as the matrix of 
second derivatives is given by the covariance matrix of constraints
\beq
\label{secondDer}
\frac{\partial^2 \mathcal{ L'}}{\partial \lambda_{ik} \lambda_{jl}} = 
 \la F_{ik} F_{jl} \ra - \la F_{ik} 
\ra \la F_{jl} \ra + \delta_{ij} \delta_{kl} \sigma_{ik}^2 
= cov(F_{ik}, F_{jl}) + \delta_{ij} \delta_{kl} \sigma_{ik}^2
\eeq
which is a positive-definite matrix. Therefore the 
solution is unique if it exists.

Once the Lagrange multipliers
are found, the solution of the problem is given by Eqs.(\ref{q1}) and 
(\ref{Hmn}), and the unconditional loss distribution for the bespoke portfolio is 
calculated using Eqs.(\ref{condLossDistrBespoke}) and (\ref{uncondLossDistrBespoke}).

\subsection{Pricing bespoke CDO tranches}

The analysis of the previous section applies to a given single time horizon $ T $.
To price tranches referencing a given bespoke portfolio, we need to pick a time grid
$ \{ T_i \} $. For each node on this grid, we find  
implied dichotomic index loss distributions for both reference indices 
as described above, and
then calculate the implied loss distribution $ p^B(x,T_i) $ for the bespoke as 
outlined in Sect.2.4, see Eq.(\ref{uncondLossDistrBespoke}). 
After that, tranches on the bespoke CDO are priced using standard formulae that we
provide here for completeness.

Let 
$ K_d, K_u $ be detachment/attachment points of the bespoke tranche. The term structure of 
tranche expected losses is then calculated as follows 
\beq
\label{EL}
EL_{T_i} \equiv EL_{i} = 
\frac{1}{K_u - K_d} \int dx \, p^B(x,T_i) \left[ \left( x - K_d 
\right)^{+} - \left( x - K_u \right)^{+} \right]  
\eeq
The default leg (otherwise known as contingent leg) of the tranche is given by
\beq 
\label{DL} 
\DL = N_0 \int_{0}^{T} B(0,t) dEL_t \simeq
N_0\sum_{i = 1}^{M} \frac{1}{2} \left( B_{i-1} + B_i \right) (
EL_{i} - EL_{i-1} ) 
\eeq 
where $ B(0,t)$ is a risk-free discount factor. 

The premium leg (paid by the protection buyer to the protection
seller) is given by
\bea 
\label{PL} 
\PL(S) &=& S \cdot N_0 \sum_{i = 1}^{M} \Delta_i
\left( B_i \cdot EN_{T_i} - \int_{T_{i-1}}^{T_i}  \frac{ t -
T_{i-1}}{T_i - T_{i-1}}
B(0,t) dEN_t \right) \\
&\simeq& S \cdot N_0 \sum_{i = 1}^{M} \Delta_i B_i \frac{1}{2}
\left( EN_{i-1} + EN_i \right)    
\nonumber 
\eea
where $ S $ is the tranche spread, $ \Delta_i $ is the day count
fraction, and 
\beq 
\label{It} 
EN_i \equiv EN_{T_i} =  1 - EL_{i}
\eeq
is the expected tranche outstanding notional at
time $T_i$. 
The integral term in (\ref{PL}) represents the accrued coupon due to
defaults happening  between the coupon payments dates. The integral
is calculated using the standard  approximation (see e.g. \cite{OKane})
that amounts to substitution of $ ( t -
T_{i-1})/(T_i - T_{i-1}) $ and $ B(0,t) $ by 1/2  and $ 
B_i $, respectively.
The fair (break-even) tranche par spread,  $ S $, is
determined from the par equation $ \DL = \PL(S) $. 

\subsection{Discussion}

\subsubsection{Inter-temporal consistency and time arbitrage}

The framework presented so far corresponds to 
what is known as ``single-period'' models in the literature. In this approach,
one deals only with marginal loss distributions at a given set of maturities, but not 
with joint probabilities of losses at different time horizons. Note that the knowledge
of marginal loss distributions is sufficient for pricing of CDO tranches. Reversing 
this argument, one can state that market prices of index tranches contain information
on marginal loss distributions but not on joint inter-temporal distributions, see e.g. 
\cite{AH} for a related discussion. 

In the above framework, we treat different time horizons separately from each other,
and calculate the implied distribution that matches a set of tranche expected 
losses (ELs) for each node $ T_i $ on a time grid. If the set of input tranche ELs is 
arbitrage-free across both strikes and time,
then the resulting implied distribution will be free of arbitrage across time for
loss levels corresponding to strikes in the calibration set. 
However, no-arbitrage
across time is not guaranteed in this approach for other loss levels.

While we have not encountered violations of time arbitrage in practice with 
our numerical experiments, the above implies that it may happen. Note that by 
continuity, once we
calibrate to an arbitrage-free set of tranche ELs, the time no-arbitrage holds not only
for reference strikes, but also for loss values around these strikes. Therefore, 
a simple {\it practical}
way to prevent volation of time arbitrage is to increase the number of strikes
in the calibration set.

A more principled approach to the problem of possible time arbitrage requires switching
to a dynamic framework. Similar to the way no-arbitrage across strikes is guaranteed 
by the properties of the KL-divergence (see above in Sec.2.1), no-arbitrage across time 
is ensured once we pick an arbitrage-free prior model for transition probabilities.
Such a construction will be presented in Sect.4.


\subsubsection{What form of entropy minimization should one use?}

A couple of further comments are in order here. 
First, note that Eqs.(\ref{q1}) and 
(\ref{Hmn}) imply the following form of the "true" ("posterior") joint distribution 
$ P \left(\vec{m}, \vec{X}\right) $:
\beq
\label{trueJoint}
P \left( \vec{X} , \vec{Z}_{\vec{m}} \right) = h_{\vec{m}} \prod_{i=1,2} 
P\left(\vec{X}_i | \vec{m} \right)  
= \frac{1}{Z(\lambda) } Q  \left( \vec{X}, \vec{Z}_{\vec{m}} \right) 
e^{\sum_{i,k} \lambda_{ik} \left( F_{ik} (\vec{X})
- EL_{ik} \right) }
\eeq
which could equivalently be obtained by a direct minimization of the KL-divergence 
in the form (\ref{KL1}), subject to constraints (\ref{constraints}). This would provide
a much shorter derivation of our final result (\ref{q1}), (\ref{Hmn}).
This is required, of course, for self-consistency of the method. 

Second, the lengthy derivation leading to our equations (\ref{q1}), (\ref{Hmn}) was given above with the intent to illustrate the modelling freedom in the implied loss approaches
which have recently become popular in the literature on credit portfolio modeling. Assume for the moment that instead of independent variation of 
(\ref{KL2}) with respect to both market factor distribution $ h_{\vec{m}}$ and 
conditional loss distributions $ P_i \left( \vec{X}_i | \vec{m} \right) $, we would 
fix the functional form of, say, the latter, and only optimize with respect to the 
former (see e.g. in Rosen and Saunders \cite{RS}).  It is clear from the above derivation
that such a procedure would yield a sub-optimal (in the sense of KL-divergence)
solution for the joint distribution of the market factor and sub-portfolio losses.

The latter point is easy to illustrate a bit more formally.  Assume, for the sake of argument, that we calibrate to a set of tranche expected losses for synthetic 
first loss (equity) tranches with payoff functions $ F_{ik}(x) = \min \left(x , K_k \right) $.
Assume that we fix the conditional 
loss distributions to their prior form,   $ P_i \left( \vec{X}_i | \vec{m} \right) = 
 Q_i \left( \vec{X}_i | \vec{m} \right) $. In this case, minimization of the Lagrangian 
 (\ref{Lagrange}) with respect to $ h_{\vec{m}} $ yields
 \beq
 \label{conditionalMinimization}
 h_{\vec{m}} = \frac{1}{\hat{Z}(\lambda)} g_{\vec{m}} e^{ - \sum_{ik} \lambda_{ik}
 \left( \int  d \vec{X}_i Q_i \left(\vec{X}_i | \vec{m} \right) F_{ik}( \vec{X}_i ) - EL_{ik} \right)}
 \eeq
 where $ \hat{Z}(\lambda) $ is a normalization factor, and Lagrange 
multipliers $ \lambda_{ik} $ minimize the Lagrangian function
 \beq
 \label{Lagr4}
 \hat{\mathcal{L}}''  \equiv \log \hat{Z}(\lambda) = 
\log \left( \sum_{\vec{m}} g_{\vec{m}} \prod_{i=1,2} e^{ - \sum_{k} \lambda_{ik}
 \left( \int d \vec{X}_i Q_i \left(\vec{X}_i | \vec{m} \right) F_{ik}( \vec{X}_i ) 
- EL_{ik} \right)}
 \right) 
 \eeq
Comparing this Lagrangian to (\ref{mini}), we note that in the limit 
$ \sigma_i \rightarrow 0 $, 
 they have the same gradients and hence reach their minima at the 
same point $ \vec{\lambda} = \vec{\lambda}_{\star} $. On the other hand, as long 
as functions $ e^{- \lambda_{ik} F_{ik}(x)} $ are convex in $ x $, by Jensen's 
inequality we have, for any fixed vector  $ \vec{\lambda} $:
\beq
\label{Jensen}
\int d \vec{X}_i Q \left( \vec{X}_i | \vec{m} \right) 
 e^{ - \sum_{ik} \lambda_{ik}  F_{ik}( \vec{X}_i )} 
 \leq
e^{ - \sum_{ik} \lambda_{ik} \int 
 d \vec{X}_i Q_i \left(\vec{X}_i | \vec{m} \right) F_{ik}( \vec{X}_i )}  
 \eeq
which implies that the minimum of $ \mathcal{L}' $ lies lower than the minimum of $ 
\mathcal{L}'' $. This exactly means sub-optimality of the solution (\ref{conditionalMinimization}) in terms of KL-distance between the "true" and 
"prior" joint distributions of $ \left( \vec{X}_i, \vec{m} \right) $. 

Finally, note that even though it might appear that allowing for independent variations of 
$ h_{\vec{m}} $ and $ P \left( \vec{X}_i | \vec{m} \right) $ brings "too much 
flexibility" to the problem, this is not true: adjustments to both  $ h_{\vec{m}} $ 
and $ P \left( \vec{X}_i | \vec{m} \right) $ are driven by the same set of Lagrange 
multipliers, and 
are equivalent to a single adjustment of the joint probability 
$ P \left(\vec{Z}_{\vec{m}}, \vec{X}\right) $, see (\ref{trueJoint}). In the 
nomenclature of factor models, 
this means that the least biased distortion of a prior
joint distribution $ Q \left(\vec{X}, \vec{Z}_{\vec{m}} \right) $ is 
obtained when {\it both}
the market factor distribution and conditional loss distribution are allowed to vary. 
The joint optimization of distribution $ Q \left(\vec{X}, \vec{Z} \right) $
has the same complexity as optimization of the market factor distribution alone.

\section{IMFM:  a dynamic multi-period formulation}

Here we present a dynamic generalization of the formalism of the previous section to a 
multi-period setting. We consider a (sufficiently dense) set of reference maturities
$ T_0, T_1, \ldots, T_{N_T-1} $\footnote{In practice, 
maturities $ \{ T_n \} $ may be taken e.g. with annual steps.}. We set   
 $ T_{-1} = 0 $ to be today's time. Let 
\beq
\label{vectorLoss}
\vec{X}_{i}^{n} = \left( X_{i1}(T_n), X_{i2}(T_n) \right) \; , \; \; i = 1,2
\; , \; n = 0,\ldots, N_T - 1
\eeq
be the loss in the $ i $-th portfolio at time $ t = T_n $, expressed as a tuple of losses in 
the "relevant" and complement sub-portfolios at the same time, respectively. 
Alternatively, we use
a dynamic 4D vector
\beq
\label{Lvec}
\vec{X}^{n} = \left( \vec{X}_{1}(T_n), \vec{X}_{2}(T_n) \right)
\eeq
to specify the joint sub-portfolio losses in the two index portfolios at time $ T_n $.

We assume a dynamic framework for the 2D common market factor $ 
\vec{Z}_{t}  = \left(Z_{1}(t), Z_{2}(t) \right) $, where the (time-dependent) 
components $  Z_{1}(t), Z_{2}(t) $ take values on the same 
grid (\ref{grid}). Furthermore, we assume 
that $ \vec{Z}_t $ is a right-continuous
jump process with jumps allowed only at times $ \{ T_n \} $:
\beq
\label{2dmarketVector}
\vec{Z}_t = \vec{Z}^{n} = 
\left( Z_{1}(T_n), Z_{2}(T_n) \right) \; , \; 
t \in [T_{n-1}, T_{n} [ \; , \; n = 0,1, \ldots, N_T-1
\eeq
Similar to (\ref{hn}), we introduce the time-dependent integer-valued 2D vector
$ \vec{m}^{n} $ that labels discrete states of the market factor components
on the interval $ [T_{n-1}, T_{n} [ $: 
\beq
\label{hn2}
h_{\vec{m}^{n}} \equiv P \left( \vec{Z}(t) = \vec{Z}_{\vec{m}^n} \right) 
=  P \left( Z_1(t) = z_{m_1}, Z_2(t) = z_{m_2} \right) \; , \; 
t \in [T_{n}, T_{n+1} [ \; , \; n = 0,1, \ldots, N_T -1   
\eeq 
We assume a Markovian setting in our model, i.e. the state variables at time $ T_{n}$ 
depend only on state variables at previous time $ T_{n-1} $ but not on their prior history:
\beq
\label{Markov}
P \left( \vec{X}^{n} , \vec{m}^{n} | \vec{X}^{0} \ldots, \vec{X}^{n-1} , 
\vec{m}^{0}, \ldots, \vec{m}^{n-1} \right) = 
P \left( \vec{X}^{n} , \vec{m}^{n} | \vec{X}^{n-1}, \vec{m}^{n-1} \right)
\eeq
We now construct a multi-period generalization of the single-period setting of the previous section, where one infers transition probabilities $ 
P \left( \vec{X}^{n} , \vec{m}^{n} | \vec{X}^{n-1}, \vec{m}^{n-1} \right) $ instead of 
marginal probabilities $  P \left( \vec{X}^{n} , \vec{m}^{n}  \right) $. The model is 
constructed (equivalently, calibrated) within a bootstrap procedure in the time 
dimension, starting with the first maturity $ T_0 $.
In treating the time dimension of the problem, our approach is 
similar to that developed in Ref.\cite{IH2002} in a somewhat different context of 
pricing equity derivatives. 

\subsection{The first maturity $ T_0 $}

For the first maturity $ T_0 $, we assume that today
($ t = 0 $, or equivalently $ n = -1 $) we have no losses in either portfolio, and the value of
the market factor prior to time $ T_{-1}  $ is fixed somehow (this value is 
irrelevant for calculation of the joint law of $ \vec{X}(t), \vec{m}(t) $ at time $ t = T_0 $).
The marginal joint distribution $ P \left(\vec{X}^{0}, \vec{m}^{0} \right) $ is 
then calculated using the formalism of the previous section, see Eq.(\ref{trueJoint}):
\bea
\label{firstMat}
P \left(\vec{X}^{0} , \vec{Z}_{\vec{m}^{0}} \right) &=& h_{\vec{m}^{0}} \prod_{i=1,2} P_i 
\left(\vec{X}_i^{0} | \vec{m}^{0} \right)  = \frac{1}{Z_0(\lambda) } 
Q \left(\vec{X}^{0}, \vec{Z}_{\vec{m}^{0}} \right) 
e^{\sum_{i,k} \lambda_{ik} \left( F_{ik} (\vec{X}^{0})
- EL_{ik}^{0} \right) } \nonumber \\
Z_0 ( \lambda) &=& \sum_{\vec{m}^{0}} g_{\vec{m}^{0}}  \int d \vec{X}^{0} 
Q \left( \vec{X}^{0}, \vec{Z}_{\vec{m}^{0}} \right) 
e^{\sum_{i,k} \lambda_{ik} \left( F_{ik} (\vec{X}^{0})
- EL_{ik}^{0} \right) } 
\eea
where $ Q( \left(\vec{Z}_{\vec{m}^{0}}, \vec{X}^{0} \right) $ is the prior joint distribution 
of the state variables at time $ T_0 $.
  
\subsection{Further maturities $ T_n $, $ n = 1, \ldots, N_T $}

Now we assume that we have 
calculated the implied joint distribution for the previous maturity $ T_{n-1} $, 
and we want to move one step on the time grid, i.e.
calculate the transition probabilities 
$ P \left( \vec{X}^{n} , \vec{m}^{n} | \vec{X}^{n-1}, \vec{m}^{n-1} \right) $. The latter 
should be calculated (calibrated) in a way that ensures consistency with pricing 
data for maturity $ T_1 $ as seen {\it today}, at time $ t = 0 $. 

Our method to calculate transition probabilities 
$ P \left( \vec{X}^{n} , \vec{m}^{n} | \vec{X}^{n-1}, \vec{m}^{n-1} \right) $ is based on minimizing of a suitable KL-divergence, similar to the procedure used above 
in the single-period setting. The only difference from the previous case is that now 
we have to minimize the {\it conditional} cross entropy of transition probabilities
$ P \left( \vec{X}^{n} , \vec{m}^{n} | \vec{X}^{n-1}, \vec{m}^{n-1} \right) $ 
and $ Q \left( \vec{X}^{n} , \vec{m}^{n} | \vec{X}^{n-1}, \vec{m}^{n-1} \right) $.
The latter is defined as follows (see e.g. \cite{CT}):
\beq
\label{relTransEnt}
D \left[ P || Q \right] 
= \sum_{\vec{m}^{n-1}, \vec{m}^{n}} \int d \vec{X}^{n-1} d \vec{X}^{n}  
P \left(\vec{X}^{n}, \vec{m}^{n} , \vec{X}^{n-1}, \vec{m}^{n-1} \right)  
\log \frac{ P \left(\vec{X}^{n}, \vec{m}^{n} |  \vec{X}^{n-1}, \vec{m}^{n-1} \right) }{
Q \left( \vec{X}^{n}, \vec{m}^{n} |  \vec{X}^{n-1}, \vec{m}^{n-1} \right) }
\eeq
which can be interpreted as the KL-divergence of transition probabilities 
$ P \left( \vec{X}^{n} , \vec{m}^{n} | \vec{X}^{n-1}, \vec{m}^{n-1} \right) $ 
and $ Q \left( \vec{X}^{n} , \vec{m}^{n} | \vec{X}^{n-1}, \vec{m}^{n-1} \right) $ viewed 
as a function of initial values $ \vec{X}^{n-1} , \vec{m}^{n-1} $, and averaged  
over these initial values  using the marginal
joint probability distribution $  P \left( \vec{X}^{n-1}, \vec{m}^{n-1} \right)  $ 
calculated at the previous time step:
\beq
\label{KL3}
D \left[  P || Q \right] 
= \sum_{\vec{m}^{n-1}} \int d \vec{X}^{n-1} P  \left( 
\vec{X}^{n-1}, \vec{m}^{n-1} \right) 
 D( \vec{X}^{n-1}, \vec{m}^{n-1} ) 
\eeq
Introducing transition probabilities for the market factors
\beq
\label{transitionMarket}
h \left( \vec{m}^n | \vec{m}^{n-1}, \vec{X}^{n-1} \right) = 
P \left[ \vec{Z}^{n} = \vec{Z}_{\vec{m}^n} | \vec{Z}^{n-1} = 
\vec{Z}_{\vec{m}^{n-1}} , \vec{X}^{n-1} \right]
\eeq
and the corresponding prior probabilities
\beq
\label{transitionMarketPrior}
g \left( \vec{m}^n | \vec{m}^{n-1}, \vec{X}^{n-1} \right) = 
P^{(prior)} \left[ \vec{Z}^{n} = \vec{Z}_{\vec{m}^n} | \vec{Z}^{n-1} = 
\vec{Z}_{\vec{m}^{n-1}} , \vec{X}^{n-1} \right]
\eeq
the conditional KL-divergence (\ref{KL3}) can be put in a more suggestive form
(compare with (\ref{KL2})):
\bea
\label{KL4}
D \left[  P || Q \right] 
&=& \sum_{\vec{m}^{n-1}} \int d \vec{X}^{n-1} P  \left[\vec{X}^{n-1}, \vec{m}^{n-1} \right] 
\left(  \sum_{\vec{m}^{n}} h \left( \vec{m}^n | \vec{m}^{n-1}, \vec{X}^{n-1} \right)
 \log \frac{ h \left( \vec{m}^n | \vec{m}^{n-1}, \vec{X}^{n-1} \right)}{
 g \left( \vec{m}^n | \vec{m}^{n-1}, \vec{X}^{n-1} \right) }  \right. \nonumber \\
 &+&  \left. \int d \vec{X}^{n}  
 P \left( \vec{X}^{n}|  \vec{m}^{n} , \vec{X}^{n-1}, \vec{m}^{n-1} \right) 
 \log \frac{  P \left( \vec{X}^{n}|  \vec{m}^{n} , \vec{X}^{n-1}, \vec{m}^{n-1} \right) }{
  Q \left( \vec{X}^{n}|  \vec{m}^{n} , \vec{X}^{n-1}, \vec{m}^{n-1} \right) } \right) 
\eea
Our pricing constraints now take the following form:
\beq
\label{constraints2}
\sum_{\vec{m}^{n-1}} \int d \vec{X}^{n-1}  P  \left( 
\vec{X}^{n-1}, \vec{m}^{n-1} \right)  
\sum_{\vec{m}^{n} }
\int d \vec{X}^{n}  P  \left( 
\vec{X}^{n}, \vec{m}^{n} | \vec{X}^{n-1}, \vec{m}^{n-1}  \right)
F_{ik} (\vec{X}^{n}) = EL_{ik}^{n}
\eeq
While the problem of minimization of (\ref{KL4}) subject to constraints (\ref{constraints2})
admit more general priors, in what follows we 
restrict ourselves to a particular sort of priors. First we assume that 
prior transition probabilities for 
the market factor depend only on the previous value of the market factor but not on the 
previous loss level:
\beq
\label{simplePriorH}
g \left( \vec{m}^n | \vec{m}^{n-1}, \vec{X}^{n-1} \right)  = 
g \left( \vec{m}^n | \vec{m}^{n-1} \right) 
\eeq
A possible candidate for the prior market factor transition matrix can be 
a simple birth-and-death discrete-time Markov chain.

Second, we assume that the probability of reaching loss $ \vec{X}^{n} $ by the end of 
period $ [T_{n-1}, T_n [ $ depends only on previous losses $ \vec{X}^{n-1} $ 
and the value  $ \vec{m}^{n} $ of the market factor on this interval, but not 
on its previous value:
\beq
\label{simplePriorQ}
Q_i  \left( \vec{X}_i^{n} |  \vec{m}^{n}, \vec{X}_i^{n-1}, \vec{m}^{n-1}  \right)    
= 
Q_i \left( \vec{X}_i^{n} |  \vec{m}^{n}, \vec{X}_i^{n-1}  \right)
\eeq	
Within such reduced class of admissible priors, for the latter we can use simple 
single-period models such as a two-factor Gaussian copula (or the RFL model) that
we used above in a static setting of Sect.3.

A brief inspection of Eqs.(\ref{KL4}) and (\ref{constraints2}) establishes that the 
results of a corresponding Lagrangian optimization problem of 
minimization of (\ref{KL4}) subject to constraints (\ref{constraints2}) can be 
immediately read off the previous 
formulae for a single-period case, provided all marginal probabilities should substituted 
by transition probabilities of transition from the previous state $ \left( \vec{X}^{n-1}, 
\vec{Z}^{n-1} \right) $.
This observation allows one to immediately write down the result
(which can of course be readily checked following steps similar to those described 
above for the single-period setting):
\bea
\label{answerMulti2}
P_i  \left(\vec{X}_i^{n} |  \vec{m}^{n}, \vec{X}_i^{n-1} \right) &=&
\frac{1}{Z_{i \lambda}^k \left( \vec{m}^{n} , \vec{X}_i^{n-1}\right)}
Q_i \left( \vec{X}_i^{n} |  \vec{m}^{n}, \vec{X}_i^{n-1}  \right)
e^{ \sum_{k} \lambda_{ik} \left( F_{ik}(\vec{X}_i^{n}) - EL_{ik}^{n} \right) }
\nonumber \\
Z_{i \lambda}^{n} \left( \vec{m}^{n}, \vec{X}_i^{n-1}  \right) &=& 
 \int d \vec{X}_i^{n}
Q_i \left( \vec{X}_i^{n} |  \vec{m}^{n}, \vec{X}_i^{n-1}  \right) 
e^{ \sum_{k} \lambda_{ik} \left( F_{ik}(\vec{X}_i^{n}) - EL_{ik}^{n} \right) } \\
h \left( \vec{m}^n | \vec{m}^{n-1}, \vec{X}^{n-1} \right)  
 &=& \frac{1}{\hat{Z}_{\lambda}^{n} 
\left(\vec{X}^{n-1}, \vec{m}^{n-1} \right) }
g \left( \vec{m}^n | \vec{m}^{n-1} \right)  
 \prod_{i=1,2}
Z_{i \lambda}^{n} \left( \vec{m}^{n} , \vec{X}_i^{n-1}  \right) \nonumber \\
\hat{Z}_{\lambda}^{n} 
\left( \vec{m}^{n-1} , \vec{X}^{n-1}\right) & = & 
\sum_{\vec{m}^{n}} 
g \left( \vec{m}^n | \vec{m}^{n-1} \right)
Z_{1 \lambda}^{n} \left( \vec{X}_1^{n-1} , \vec{m}^{n}   \right) 
Z_{2 \lambda}^{n} \left( \vec{X}_2^{n-1} , \vec{m}^{n}   \right)
\nonumber 
\eea 
Note that the first of Eqs.(\ref{answerMulti2}) shows that our framework is 
arbitrage-free as long as the prior model is, irrespective of the model 
$ g \left( \vec{m}^n | \vec{m}^{n-1} \right)  $ for the dynamic market factor.
Furthermore, the third of Eqs.(\ref{answerMulti2}) demonstrates that our model
produces credit contagion: even though we started with a 
prior transition probabilities $ g \left( \vec{m}^n | \vec{m}^{n-1} \right) $ that
were independent of the previous loss levels $ \vec{X}_i^{n-1} $, the 
``true'' transition probabilities 
$ h \left( \vec{m}^n | \vec{m}^{n-1}, \vec{X}^{n-1} \right) $ do depend on them.
Finally, note that Eqs.(\ref{answerMulti2}) can also be combined into a 
formula for joint transition probabilities
\beq
\label{posteriorJoint}
P \left( \vec{m}^{n}, \vec{X}^{n} | \vec{m}^{n-1}, \vec{X}^{n-1} \right) 
= \frac{1}{\hat{Z}_{\lambda}^{n} 
\left( \vec{m}^{n-1} , \vec{X}^{n-1}\right)}
Q \left( \vec{m}^{n}, \vec{X}^{n} | \vec{m}^{n-1}, \vec{X}^{n-1} \right)
e^{ \sum_{i,k} \lambda_{ik} \left( F_{ik}(\vec{X}_i^{n}) - EL_{ik}^{n} \right) }
\eeq
Here the values of Lagrange multipliers $ \lambda_{ik} $ should be substituted by the 
solution of a  $ (N_1 + N_2 + 4) $-dimensional convex optimization
problem with the Lagrangian function (compare with Eq.(\ref{mini}))
\beq
\label{optimization2}
\mathcal{L} = \sum_{\vec{m}^{n-1}} \int d \vec{X}^{n}  P 
\left( \vec{m}^{n-1}, \vec{X}^{n-1} \right) \log 
\hat{Z}_{\lambda}^{n} 
\left( \vec{m}^{n-1}, \vec{X}^{n-1}  \right) 
+ \frac{1}{2} \sum_{ik} \lambda_{ik}^2 \sigma_{ik}^2
\eeq 

\subsection{Completing one step of time bootstrap}

Having computed the transition probability $ P \left[ \vec{X}^{n} , \vec{m}^{n} | \vec{X}^{n-1}, \vec{m}^{n-1} \right] $ and using the marginal probability 
$ P \left[\vec{X}^{n-1}, \vec{m}^{n-1} \right] $ known from the previous time step, we know 
use them to calculate the marginal joint distribution of state variables at time $ T_k $:
\beq
\label{newMarginal}
 P \left( \vec{m}^{n} , \vec{X}^{n}   \right) =  
\sum_{\vec{m}^{n-1}} \int d \vec{X}^{n-1} 
  P \left( \vec{m}^{n-1}, \vec{X}^{n-1} \right) 
P \left( \vec{m}^{n}, \vec{X}^{n} | \vec{m}^{n-1}, \vec{X}^{n-1} \right)
 \eeq
 Unless we reached the final maturity $ T_{N_k} $ on the time grid, we know increment 
 $ k \rightarrow k + 1 $, coming up with a problem identical to that just solved for the 
 previous period, where (\ref{newMarginal}) plays the role of the marginal joint distribution from the previous step.
 
\section{Generalizations}

\subsection{More market factors}

Formally generalization to more than two market factors is straightforward and follows
the same lines as developed above. However, for higher number of factors (three and more?)
Monte Carlo may be preferred for integration over market factor distribution as well
as for calculation of conditional loss distributions.

\subsection{More index chunks in the bespoke}

Assume that we have to price a tranche on a bespoke composed of three chunks belonging
to three different indices $ \{I_i\}_{i=1}^{3} $. 
Such calculation would proceed along the same line as described above, with a 
simultaneous calibration to tranches referencing all indices $ I_i $, while the 
conditional loss distribution (\ref{condLossDistrBespoke}) would now be given by
a two-dimensional convolution instead of a one-dimensional one. Note that for this case 
a three-factor version of the model may be preferred to a two-factor one.

\subsection{Bespoke portfolios with ``bespoke'' names}

So far, we have assumed that each name in a bespoke portfolio belongs to some credit 
index. In practice, this is often not the case, as majority of actual bespoke portfolios
typically have some ``bespoke'' names, i.e. names that do not belong to any credit index.

Let us assume that we are given a bespoke portfolio of the following form:
$ \Pi_B = \Pi_{11} \cup \tilde{\Pi}_{21} $, where $ \Pi_{11} $ stands for a sub-set of 
names from an index portfolio $ I_1 $ (e.g. CDX.NA.IG) and $ \tilde{\Pi}_{21} $ is 
a sub-set of ``bespoke'' names that do not belong in any credit index. 
For example, the sub-set $ \tilde{\Pi}_{21} $ can be composed of 
non-investment grade US names. In this case, we would take the CDX.NA.HY portfolio
is the closest second benchmark security to our bespoke set.
 
More generally, assume that
names in $ \tilde{\Pi}_{21} $ are similar in terms of their geographic sectors 
and ratings to names in another index $ I_2 $. We can then pose (and solve) the problem
of finding a sub-set $ \Pi_{21} $ of names from $ I_2 $ that provides the best 
approximation of the set  $ \tilde{\Pi}_{21} $ in terms of sector diversification,
average expected loss, and possibly higher moments of the loss distribution.
This is a combinatorial optimization problem that can be solved e.g. using a genetic 
algorithm (details will be given elsewhere).
Once such a proxy index sub-portfolio is found,   
our problem of pricing bespoke tranches can be reduced to the previous problem
of pricing tranches on a $ I2$-type bespoke portfolio. 

The approximating sub-portfolio $ \Pi_{21} $ is generally expected to have an average
expected loss (EL) that is not exactly equal to the EL of the original set  
$ \tilde{\Pi}_{21} $. This could lead to a mispricing of bespoke tranches. However,
this can be easily cured by adjusting the loss distribution found for the sub-portfolio
$ \Pi_{21} $ to match the expected loss of $ \tilde{\Pi}_{21} $. To this end, we use 
the MCE method once again. 

Let $ P \left(X | \vec{m} \right) $ be the ``true'' conditional
loss distribution of the set $ \tilde{\Pi}_{21} $, and $ EL $ be its total expected loss.
Further let $ Q \left(X | \vec{m} \right) $ and $ \{ h_{\vec{m}} \} $ be the 
conditional loss distribution of the approximating sub-portfolio $  \Pi_{21} $ and
the implied market factor distribution, respectively. We assume that 
$ Q \left(X | \vec{m} \right) $ and $ \{ h_{\vec{m}} \} $ are calculated as described 
in Sect.3. We assume that the net difference between loss distribution of sets 
$ \tilde{\Pi}_{21} $ and $ \Pi_{21} $ amounts to an adjustment of the conditional loss 
distribution, while keeping the market factor distribution intact. This leads to a 
variational problem with the following Lagrangian function
\beq
\label{LagrangeAdj}
\mathcal{L}  =     
\sum_{\vec{m}} h_{\vec{m}}  \int d X \, P
\left( X | \vec{m} \right)
\log \frac{ P \left(X | \vec{m} \right) }{ Q \left(X | \vec{m} \right)}  
+  \lambda  
\left( \sum_{\vec{m}} h_{\vec{m}} \int d X \, X \,
P \left(X | \vec{m} \right)   - EL \right) 
\eeq 
where $ \lambda $ is a Lagrange multiplier enforcing the expected loss constraint.
Minimizing this wrt $ P \left( X | \vec{m} \right) $, we obtain
\beq
\label{adjustCondDistr}
 P \left( X | \vec{m} \right) = \frac{1}{Z_{\vec{m}} 
(\lambda)} Q \left(X | \vec{m} \right)
e^{ -\lambda( X - EL )}  
\eeq
where $ Z_{\vec{m}}(\lambda) $ is a normalization factor:
\beq
\label{normZ}
Z_{\vec{m}}(\lambda) = \int d X \,  Q \left(X | \vec{m} \right)
e^{ -\lambda( X - EL )}
\eeq
By substituting (\ref{adjustCondDistr}) back into (\ref{LagrangeAdj})
and changing the sign $ \mathcal{L} \rightarrow - \mathcal{L} $ , we obtain 
\beq
\label{lambdaOpt}
\mathcal{L}  =  \sum_{\vec{m}} h_{\vec{m}} \log Z_{\vec{m}}(\lambda)
\eeq
The value of $ \lambda $ should be found by minimization of (\ref{lambdaOpt}). As 
before, this is a convex
problem, therefore it has a unique solution that can be obtained in about a second.

To summarize, the procedure of calibrating a model for a bespoke portfolio that has 
both names from an index portfolio $ I_1 $ and names {\it similar} (but not identical) 
to some names from
an index portfolio $ I_2 $ amounts to three separate and consecutive optimizations:
\begin{itemize}
\item Find an optimal sub-portfolio $ \Pi_{21} $ of portfolio $ I_2 $ that 
approximates the ``bespoke''
set of names $ \tilde{\Pi}_{21} $ in the bespoke portfolio 
\item Use the MCE procedure of Sect.4 to calculate implied loss distributions in 
index sub-portfolios
\item Perform one-dimensional minimization of function (\ref{lambdaOpt}) to find 
the optimal value $ \lambda = \lambda^{*}$. This value is then used in 
(\ref{adjustCondDistr}) to calculate an adjusted conditional loss distribution 
for the approximating sub-portfolio.
\end{itemize}
After that, bespoke tranches can be priced following steps described in Sect.2.4.
   
\section{Numerical examples}

Here we illustrate performance of the model and the difference of the resulting prices
from those obtained with the Base Correlation method using two sets of artificial 
bespoke portfolios, both made of names that belong in index portfolios. We will only 
present results obtained with a pseudo-dynamic version of the model described in Sect.3,
results obtained with a dynamic version will be reported elsewhere.

\begin{figure}
\begin{center}
\includegraphics[width=10cm, height=130mm]{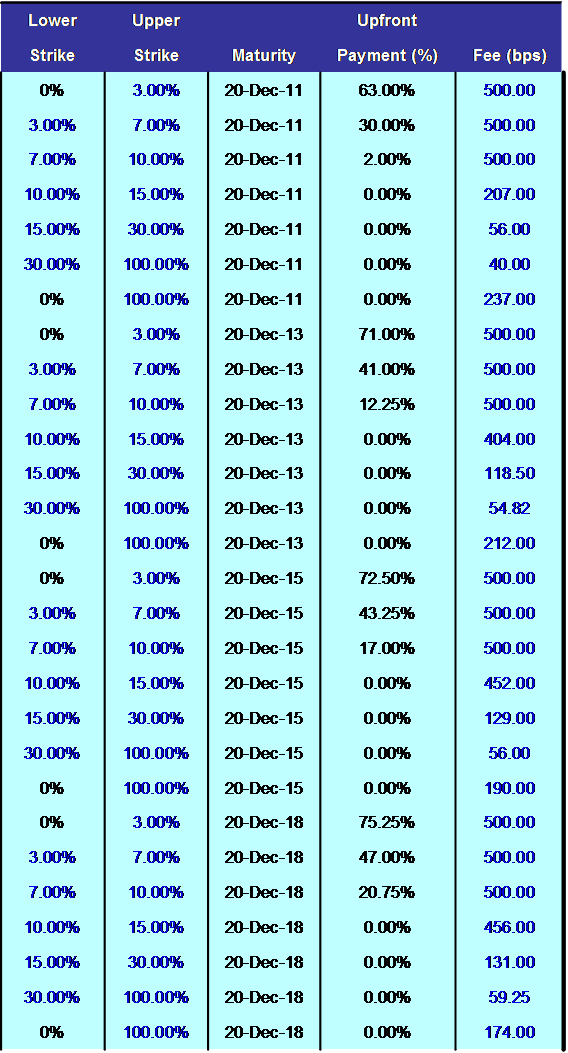}
\caption{Tranche quotes for CDX.NA.IG11 on 04/15/09.} 
\label{fig:CDX_1}
\end{center}
\end{figure}

\begin{figure}
\begin{center}
\includegraphics[width=10cm, height=130mm]{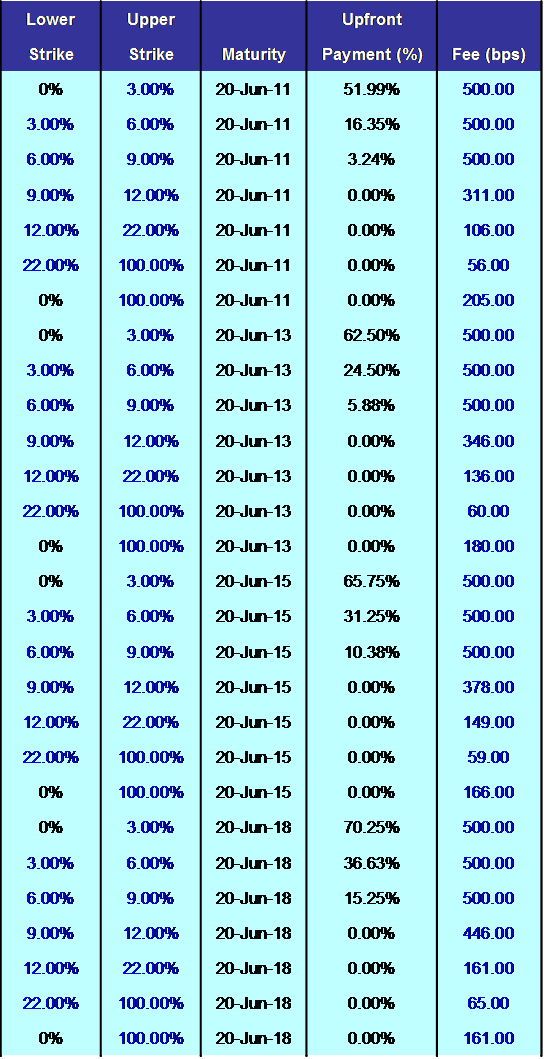}
\caption{Tranche quotes for iTraxx Europe S9 on 04/15/09.} 
\label{fig:iTraxx}
\end{center}
\end{figure}

Our first example deals with a bespoke portfolio 
composed of 50 highest spread names from
CDX.NA.IG11 and 50 highest spread names from iTraxx Europe 9 index portfolios.
The pricing date is 04/15/2009. Quotes for tranches on CDX.NA.IG11 and
iTraxx Europe S9 indices are shown in Fig.~\ref{fig:CDX_1} and Fig.~\ref{fig:iTraxx}, 
respectively.

\begin{figure}
\begin{center}
\includegraphics[width=10cm, height=70mm]{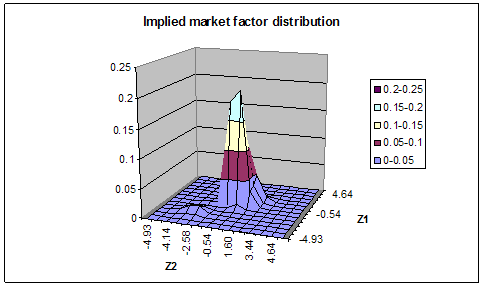}
\caption{Implied distribution of the market factor on 04/15/09, calibration to CDX.NA.IG11 and
iTraxx Europe S9.} 
\label{fig:MarketFactorDist_CDX_iTraxx}
\end{center}
\end{figure}

We calibrate the model to a set of tranche expected losses obtained using BSLP
model of Ref.\cite{AH} (Results obtained using calibration to base correlation 
quotes will be reported below.).
Calibration results are shown in Tables~\ref{calib_CDX} and 
~\ref{calib_iTraxx}, where we display relative errors between the calculated
and input tranche ELs obtained in the joint calibration of the model to CDX.NA.IG11 and 
iTraxx Europe S9 data. Note that as losses in senior tranches at short maturities are 
very small, large relative errors observed for these tranches  
give rise to very small pricing error in terms of tranche
par spreads, thus rendering calibration nearly perfect. The implied distribution of the 
market factor is shown in Fig.~\ref{fig:MarketFactorDist_CDX_iTraxx}.

\begin{table}
\vspace{5mm}
\begin{tabular}{|c|c|c|c|c|c|c|c|c|c|c|c|c|c|c|}
\hline
Tranche & 0-3\% & 3-7\% & 7-10\% & 10-15\% & 15-30\% & $ \Pi_{11} $ & 
$ \Pi_{12} $ \\
\hline
1Y & -0.010 & -0.067 & 71.313 & -90.575 & -81.617 & 0.016 & -0.038 \\
2Y & -0.005 & -0.011 & - 0.024 & - 0.040 & - 0.468 & 0.005 & - 0.004 \\
3Y & -0.000 & - 0.002 & 0.000 & 0.003 & 0.003 & 0.000 & -0.001 \\
4Y & 0.0059 & 0.002 & 0.0163 & 0.006 & 0.018 & - 0.002 & - 0.000 \\
5Y & 0.007  & 0.002 & 0.013 & 0.005 & 0.012 & - 0.001 & 0.000 \\
\hline
\end{tabular}
\caption{Relative errors $ \frac{EL_{model} - EL_{input}}{EL_{input}} $ 
(in percent) between input ELs for CDX.NA.IG11 and calculated model 
values obtained with joint calibration to CDX.NA.IG11 and iTraxx Europe 9 data on
04/15/2009.}
\label{calib_CDX} 
\end{table} 

\begin{figure}
\begin{center}
\includegraphics[width=10cm, height=70mm]{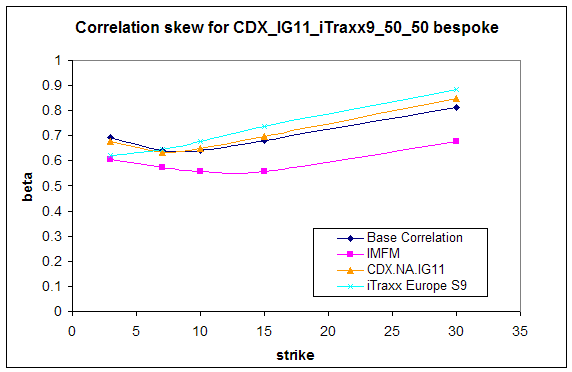}
\caption{Equivalent correlation skew for the mixed CDX IG11-iTraxx9 
bespoke portfolio priced on 04/15/09.} 
\label{Correlation_skew_CDX_IG11_iTraxx_9_50_50_besp}
\end{center}
\end{figure}

\begin{table}
\vspace{5mm}
\begin{tabular}{|c|c|c|c|c|c|c|c|c|c|c|c|c|c|c|}
\hline
Tranche & 0-3\% & 3-6\% & 6-9\% & 9-12\% & 12-22\%  & $ \Pi_{21}$ & $ \Pi_{22} $ \\
\hline
1Y & -0.013 & -0.033 & -0.063 & - 4.831 & -25.914 & 0.021  & -0.109  \\
2Y & -0.001 & 0.008  & -0.018 & 0.072   & 0.018   & -0.000 & - 8.039 \\
3Y &  0.001 & 0.006  & 0.006  & -0.004  & 0.019 & -0.001 & - 2.403 \\
4Y &  0.001 & 0.004  & 0.010  & 0.000 & 0.017 & - 0.001 & - 0.572 \\
5Y &  0.001 & 0.002 & 0.007   &0.006  & 0.012 & - 0.001 & - 0.106 \\
\hline
\end{tabular}
\caption{Relative errors  $ \frac{EL_{model} - EL_{input}}{EL_{input}} $ (in percent) 
between input ELs for  iTraxx Europe 9 and calculated model 
values obtained with joint calibration to CDX.NA.IG11 and iTraxx Europe 9 data on
04/15/2009.}
\label{calib_iTraxx} 
\end{table} 
The result of pricing tranches referencing our bespoke portfolio are shown in 
Table~\ref{CDX_IG11_iTraxx_bespoke_pricing} where strikes are chosen to coincide
with standard strikes of the CDX.NA.IG portfolio. We show the results obtained 
with two sets of correlation parameters: for Case 1, we use $ \rho = 0.5 $,
$ \alpha = 0.3 $, and for Case 2, we set $ \rho = 0.75 $, $ \alpha = 0.0 $.
Note that inter-sector correlations are decreased in the second case by 10\% relatively
to the first one, see Eq.(\ref{pairwise2}). Therefore, the effect of transition 
from Case 1 to Case 2 is as expected: junior spreads go up, while senior spreads go 
down\footnote{Note that, as we enforce portfolio loss constraints together with tranche
expected loss constraints, we do not include super-senior tranches in our calibration 
set in order 
to avoid linear dependencies between constraints (which would follow otherwise as 
the sum of all tranche losses should equal the portfolio loss). As a result, the 
impact of changed inter-sector correlation on the price of a super-senior tranche is 
less pronounced.}. On the other hand, for both choices of correlation parameters
we find that senior mezzanine and junior senior tranches are substantially more 
expensive in our model than in the Base correlation approach. We can also convert 
prices into equivalent base correlations, see 
Fig.~\ref{Correlation_skew_CDX_IG11_iTraxx_9_50_50_besp}.


\begin{table}
\begin{tabular}{|c|c|c|c|c|c|c|c|c|c|c|}    
\hline
\multicolumn{2}{c}{Tranche} & \multicolumn{3}{c}{Base correlation}
& \multicolumn{3}{c}{IMFM, case 1} & \multicolumn{3}{c}{IMFM, case 2} \\
\hline
Low & Upper  & Par  & Risky & Default  & 
Par  & Risky  & Default  & 
Par  & Risky  & Default  \\
strike & strike & spread & annuity & leg & 
spread & annuity & leg & spread & annuity & leg  \\
\hline
0\%  & 3 \% & 6000.9 & 1.435 & 0.861 & 6358.9 & 1.428 & 0.908 & 6623.1 & 1.382 & 0.915 \\
3\%  & 7\%  & 2914.1 & 2.476 & 0.722 & 3035.5 & 2.543 & 0.772 & 3116.8 & 2.535 & 0.790 \\
7\%  & 10\% & 1655.8 & 3.325 & 0.551 & 2014.6 & 3.224 & 0.649 & 1991.1 & 3.247 & 0.647 \\
10\% & 15\% & 893.6  & 3.948 & 0.353 & 1248.2 & 3.858 & 0.482 & 1229.6 & 3.835 & 0.472 \\
15\% & 30\% & 329.65 & 4.507 & 0.149 & 409.2  & 4.524 & 0.185 & 403.4  & 4.538 & 0.183 \\
30\% & 100\%& 130.15 & 4.407 & 0.057 & 63.4   & 4.786 & 0.030 & 63.6   & 4.785	& 0.031
\\  
\hline
\end{tabular}
\caption{Pricing of tranches on the mixed CDX IG11-iTraxx9 
bespoke portfolio with 5Y to maturity on 04/15/09. Case 1: $ \rho = 0.5 $,
$ \alpha = 0.3 $. Case 2: $ \rho = 0.75 $, $ \alpha = 0.0 $.} 
\label{CDX_IG11_iTraxx_bespoke_pricing}
\end{table}

To test the impact of different interpolation schemes used to calculate tranche
expected losses on the time grid, we have performed comparison of results 
obtained with calibration of IMFM to tranche ELs obtained with the Base correlation (BC)
model instead of ELs generated by BSLP. The results are shown in 
Table~\ref{IMFM_calib_BSLP_vs_BC}. It turns out that calibration to BC-generated 
tranche ELs produces larger pricing errors for short time maturities comparing to 
calibration to BSLP-generated ELs. The resulting bespoke tranche prices are however reasonably
close to numbers obtained with the former method (we skip the results to save space).
\begin{table}
\begin{tabular}{|c|c|c|c|c|c|c|c|c|c|c|}    
\hline
\multicolumn{2}{c}{Tranche} & \multicolumn{3}{c}{Base correlation}
& \multicolumn{3}{c}{IMFM, case 1} & \multicolumn{3}{c}{IMFM, case 2} \\
\hline
Low & Upper  & Par  & Risky & Default  & 
Par  & Risky  & Default  & 
Par  & Risky  & Default  \\
strike & strike & spread & annuity & leg & 
spread & annuity & leg & spread & annuity & leg  \\
\hline
0\%  & 3 \% & 6000.9 & 1.435 & 0.861 & 7465.5 & 1.211 & 0.904 & 6623.1 & 1.382 & 0.915 \\
3\%  & 7\%  & 2914.1 & 2.476 & 0.722 & 3407.8 & 2.271 & 0.774 & 3116.8 & 2.535 & 0.790 \\
7\%  & 10\% & 1655.8 & 3.325 & 0.551 & 2161.5 & 2.925 & 0.632 & 1991.1 & 3.247 & 0.647 \\
10\% & 15\% & 893.6  & 3.948 & 0.353 & 1324.5 & 3.540 & 0.469 & 1229.6 & 3.835 & 0.472 \\
15\% & 30\% & 329.65 & 4.507 & 0.149 & 433.3  & 4.435 & 0.192 & 403.4  & 4.538 & 0.183 \\
30\% & 100\%& 130.15 & 4.407 & 0.057 & 67.3   & 4.766 & 0.032 & 63.6   & 4.785	& 0.031
\\  
\hline
\end{tabular}
\caption{Pricing of tranches on the mixed CDX IG11-iTraxx9 
bespoke portfolio with 5Y to maturity on 04/15/09. Case 1: calibration to ELs 
generated by Base correlation model. 
Case 2: calibration to ELs generated by BSLP model. For both cases, $ \rho = 0.75 $ and
$ \alpha = 0 $.} 
\label{IMFM_calib_BSLP_vs_BC}
\end{table}

\begin{figure}
\begin{center}
\includegraphics[width=10cm, height=130mm]{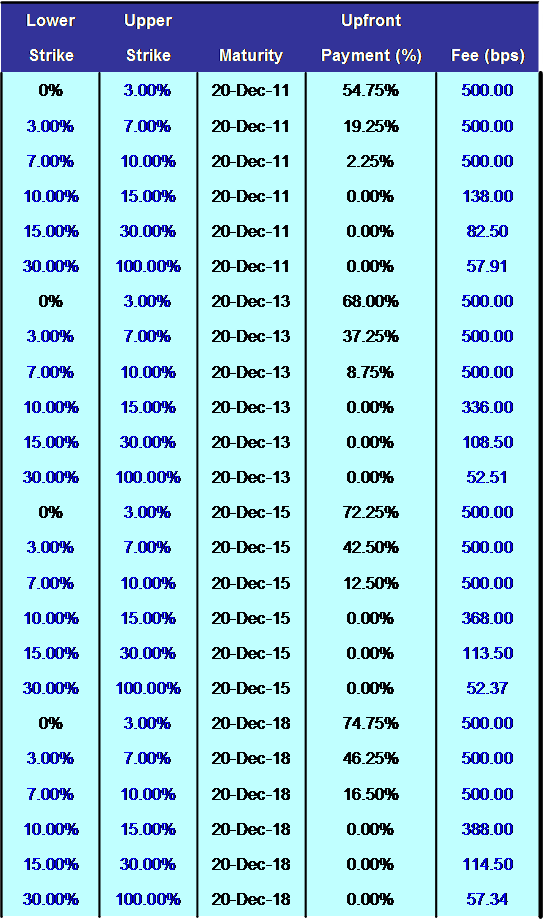}
\caption{Tranche quotes for CDX.NA.IG11 on 05/15/09.} 
\label{fig:CDX_2}
\end{center}
\end{figure}

\begin{figure}
\begin{center}
\includegraphics[width=10cm, height=130mm]{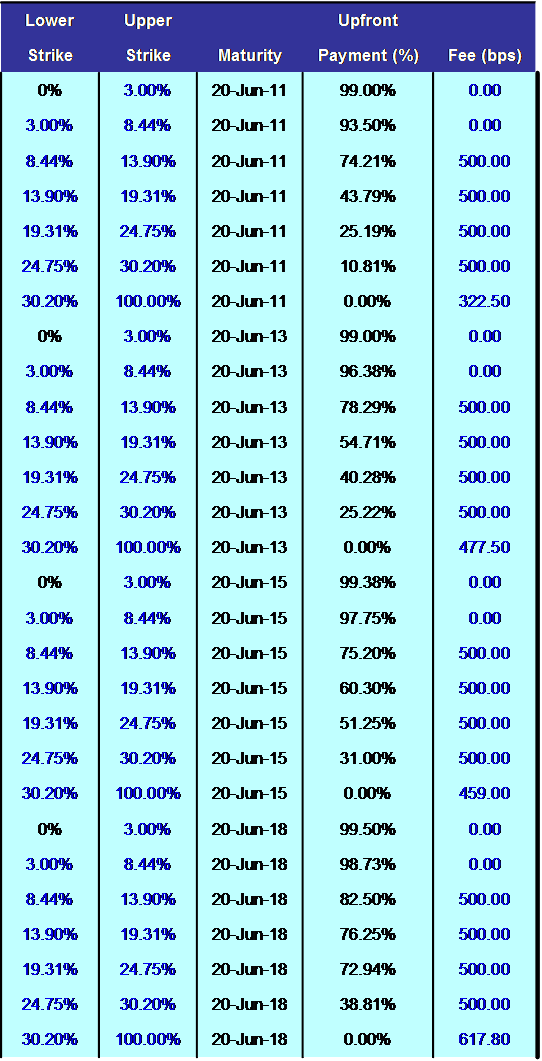}
\caption{Tranche quotes for CDX.NA.HY10 on 05/15/09.} 
\label{fig:CDX_HY}
\end{center}
\end{figure}

Our second bespoke portfolio is a mixed portfolio of 125 
US investment and non-investment grade names, priced on 05/15/09. 
It has 90 highest spread names from CDX.NA.IG11, while 35 lowest spread 
names from this index are substituted by 35 lowest spread names from CDX.NA.HY10 index.
Market quotes on index tranches are shown in Figs.~\ref{fig:CDX_2} and ~\ref{fig:CDX_HY}.

\begin{figure}
\begin{center}
\includegraphics[width=10cm, height=70mm]{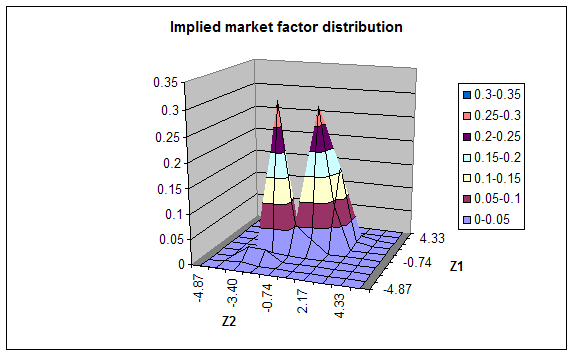}
\caption{Implied distribution of the market factor on 05/15/09, 
calibration to CDX.NA.IG11 and CDX.NA.HY10.} 
\label{fig:MarketFactorDist_CDX_IG_HY}
\end{center}
\end{figure}

Calibration results are shown in Tables~\ref{calib_CDX_2} and 
~\ref{calib_CDX_HY}, where we display relative errors between the calculated
and input tranche ELs obtained in the joint calibration of the model to CDX.NA.IG11 and 
CDX.NA.HY10 data. Similar to the above, large relative errors observed 
at short maturities do not have a material impact on the quality of calibration because
losses in senior tranches at short maturities are very small anyway.
The implied distribution of the 
market factor for this case is shown in Fig.~\ref{fig:MarketFactorDist_CDX_IG_HY}.
Unlike previous case, we now observe a pronounced bi-modal shape of the implied 
distribution.

\begin{table}
\vspace{5mm}
\begin{tabular}{|c|c|c|c|c|c|c|c|c|c|c|c|c|c|c|}
\hline
Tranche & 0-3\% & 3-7\% & 7-10\% & 10-15\% & 15-30\% & $ \Pi_{11} $ & 
$ \Pi_{12} $ \\
\hline
1Y & -0.000 & -0.005 & -0.231 & -93.745 & -84.426 & 0.000 & 0.002 \\
2Y & 0.001 & 0.000 & 0.005 & 0.030 & 0.026 & 0.001 &  0.002 \\
3Y & 0.000 & 0.000 & 0.000 & 0.000 & 0.000 & 0.000 & 0.000 \\
4Y & 0.000 & 0.000 & 0.001 & 0.001 & 0.002 & 0.000 & 0.000 \\
5Y & 0.000  & 0.000 & 0.000 & 0.000 & 0.001 & 0.000 & 0.000 \\
\hline
\end{tabular}
\caption{Relative errors $ \frac{EL_{model} - EL_{input}}{EL_{input}} $ 
(in percent) between input ELs for CDX.NA.IG11 and calculated model 
values obtained with joint calibration to CDX.NA.IG11 and CDX.NA.HY10 data on
05/15/2009.}
\label{calib_CDX_2} 
\end{table} 

\begin{table}
\vspace{5mm}
\begin{tabular}{|c|c|c|c|c|c|c|c|c|c|c|c|c|c|c|}
\hline
Tranche & 0-3\% & 3-8.4\% & 8.4-13.9\% & 13.9-19.3\% & 
19.3-24.8\% & 24.8-30.2\% & $ \Pi_{21}$ & $ \Pi_{22} $ \\
\hline
1Y & 0.000  & 0.000 & 0.000 & 0.000 & 0.000 & 0.001 & -0.001 & 0.000  \\
2Y & 0.000 & 0.000  & 0.000 & 0.000 & 0.000 & 0.000 & 0.000  & 0.000 \\
3Y &  0.001 & 0.001  & 0.001  & 0.001  & 0.001 & 0.007 & 0.001 & - 1.315 \\
4Y &  0.000 & 0.000  & 0.000  & 0.000  & 0.000 & 0.001 & 0.000 & - 5.115 \\
5Y &  0.000 & 0.000  & 0.000  & 0.000  & 0.000 & 0.001 & 0.000 & - 5.264 \\
\hline
\end{tabular}
\caption{Relative errors  $ \frac{EL_{model} - EL_{input}}{EL_{input}} $ (in percent) 
between input ELs for CDX.NA.HY10 and calculated model 
values obtained with joint calibration to CDX.NA.IG11 and CDX.NA.HY10 data on
05/15/2009.}
\label{calib_CDX_HY} 
\end{table} 
 
\begin{figure}
\begin{center}
\includegraphics[width=10cm, height=70mm]{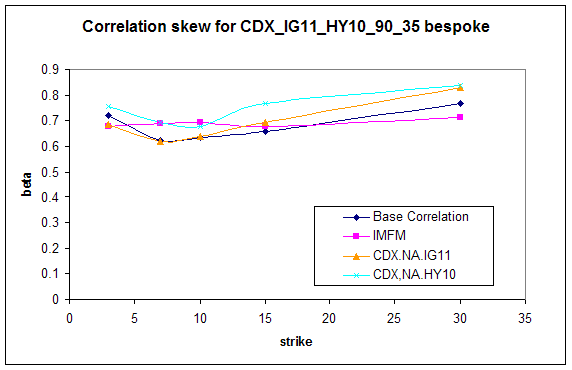}
\caption{Equivalent correlation skew for the mixed CDX IG11-CDX HY10  
bespoke portfolio priced on 05/15/09.} 
\label{Correlation_skew_CDX_IG11_HY10_9_90_35_besp}
\end{center}
\end{figure}

Results of pricing tranches on this portfolio with maturity of 5Y are shown in 
Table~\ref{CDX_IG11_HY10_bespoke_pricing}. Note that qualitatively the difference
in resulting numbers between IMFM and BC is similar to results found for our previous
example of a mixed CDX.IG-iTraxx bespoke portfolio, i.e. the largest discrepancy 
between the two models is obtained for 
senior mezzanine and junior senior tranches.  
Equivalent correlation skew in shown in 
Fig.~\ref{Correlation_skew_CDX_IG11_HY10_9_90_35_besp}.

\begin{table}
\begin{tabular}{|c|c|c|c|c|c|c|c|c|c|c|}    
\hline
\multicolumn{2}{c}{Tranche} & \multicolumn{3}{c}{Base correlation}
& \multicolumn{3}{c}{IMFM, case 1} & \multicolumn{3}{c}{IMFM, case 2} \\
\hline
Low & Upper  & Par  & Risky & Default  & 
Par  & Risky  & Default  & 
Par  & Risky  & Default  \\
strike & strike & spread & annuity & leg & 
spread & annuity & leg & spread & annuity & leg  \\
\hline
0\%  & 3 \% & 4554.1 & 1.81 & 0.821 & 3495.6 & 2.30 & 0.804 & 3503.9 & 2.29 & 0.804 \\
3\%  & 7\%  & 2449.7 & 2.87 & 0.703 & 1633.7 & 3.48 & 0.569 & 1651.9 & 3.46 & 0.572 \\
7\%  & 10\% & 1224.1 & 3.79 & 0.464 & 1105.9 & 3.80 & 0.421 & 1121.7 & 3.77 & 0.423 \\
10\% & 15\% & 738.8  & 4.25 & 0.314 & 1012.6 & 3.89 & 0.394 & 1020.1 & 3.90 & 0.398 \\
15\% & 30\% & 275.1  & 4.60 & 0.127 & 438.3  & 4.49 & 0.197 & 439.2  & 4.49 & 0.197 \\
30\% & 100\%& 72.8   & 4.54 & 0.033 & 53.7   & 4.79 & 0.026 & 52.3   & 4.79 & 0.025
\\  
\hline
\end{tabular}
\caption{Pricing of tranches on the mixed CDX IG11-CDX HY10 
bespoke portfolio with 5Y to maturity on 05/15/09. 
Case 1: $ \rho = 0.5 $,
$ \alpha = 0.3 $. Case 2: $ \rho = 0.75 $, $ \alpha = 0.0 $.
}
\label{CDX_IG11_HY10_bespoke_pricing}
\end{table}

\section{Conclusion}

In this paper we have presented the Implied Multi-Factor model (IMFM) - a semi-parametric
hybrid bottom-up/top-down model designed for pricing and risk management of tranches
on bespoke portfolios, as well as other, more exotic derivatives referencing bespoke
portfolios. The model ensures no-arbitrage in the bespoke pricing, and 
eliminates the need for 
{\it ad hoc} ``base correlation mapping rules'' that are often used by practitioners
to price bespoke CDO tranches. The alternative suggested in this paper is to use instead
standard tools of probability theory in combination with the 
Minimum Cross Entropy (MCE) method of statistical inference. The latter is 
employed to infer, an a least biased way, loss distributions of index
portfolio implied by available prices of tranches referencing these portfolios. 

Our framework differs from most of more traditional models in a number of points.
First, our approach is semi-parametric, allowing one to automatically adjust the number 
of free parameters (and calibrate them) once the size of calibration set is changed.
This produces a flexible and accurate calibration.
In addition, this approach can be easily adjusted to accomodate e.g. quotes on sub-portfolios
of index portfolios (or even tranches referencing bespoke portfolios) if traders are willing
to provide such quotes. 
Second, the model is able to control the relative importance of fitting the data versus
proximity to the ``prior'' model, thus providing means to avoid possible overfitting, or
to find a solution in sutuations where constraints cannot be satisfied exactly.
Third, calibration in our IMFM framework amounts to convex optimization in a low 
dimensional space. Therefore, the model is computationally efficient and calibrates 
within minutes rather than hours which we would expect for a more traditional bottom-up
model with a multi-factor structure. 

Last but not least, convexity of the calibration problem in our model
leads to a unique solution. 
Recall that  more traditional bottom-up approaches to
modeling credit portfolios typically give rise to objective functions having multiple
local minima. When the model is re-calibrated from one day to another, 
the presence of local minima poses a problem, as a local search algorithm 
typically used for 
calibration might end up in a different local minimum from that found a day earlier,
leading to unstable calibration and hedge ratios. 
Our model is free of such potential pitfalls. 
   
We have formulated two versions of the model. The first, single-period 
(``static'') version can be used for pricing bespoke CDO tranches. The second, 
dynamic version of the model can be used to price and risk-manage other, more exotic
portfolio derivatives, such as e.g. forward-starting tranches or tranche options.
Another potential application for the dynamic version of our model is the counterparty
risk management (including, in particular, CVA calculations) for CDO tranches.
Because the model is low-dimensional and Markovian, efficient
lattice or tree implementations are possible.

The approach presented in this paper is readily generalizable and extensible.
Even though we presented details and numerical examples only for the basic case involving
two market factors and two reference index portfolios for a given bespoke portfolio,
we have shown how to generalize this setting to more complex cases including more 
reference portfolio, more market factors, and handling names not belonging in any 
index portfolio. Further extensions including risk management of bespoke tranches and
modeling of price uncertainty of bespoke tranches due to bid/ask spreads of index 
tranches will be presented elsewhere.




\appendix
\def\theequation{\thesection.\arabic{equation}}

\def\thesection{B}	
\setcounter{equation}{0}

\section*{Appendix: Mutual information and portfolio loss partitioning}

As was mentioned above, the solution (\ref{q1}) indicates that the 
conditional independence of losses in sub-portfolios does not hold  
for the ``{\it posterior}'' model that adjust the ``prior'' distributions so that 
pricing constraints referencing portfolios as a whole are respected. In 
this section, we discuss this phenomenon in more details.

We start with an intuitive explanation. Assume we have two subportfolios $ \Pi_{11} $ 
and $ \Pi_{12} $ of an index portfolio $ \Pi_{1} $. whose losses
$ X \equiv X_{11} $ and $ Y \equiv X_{12} $ are 
assumed to be independent {\it a priori}. Obviously, if we know
the loss $ L $ in the whole portfolio with certainty, $ L = L_0 $, then 
instead of independence of 
$ X $ and $ Y $ we have a deterministic dependence between them as now $ Y = L_0 - X $.
By continuity, we should expect that when instead of knowing losses with certainty
we only know the expectation of $ L $ (as $ L$ should generally be treated as 
a random variable), we should expect a probabilistic dependence
between $ X $ and $ Y $, because the previous deterministic case would be recovered in
the limit when the loss distribution is a delta function centered at $ L_0 $:
$ \la L \ra \equiv \int dL \, L p(L) \rightarrow \int dL \, L \delta(L - L_0) = L_0 $.
This implies that the more information (less uncertainty) 
we have about the distribution of $ L $, 
the more informative $ X $ and $Y $ should be about each other. 

Let us make this intuitive argument a bit more formal.     
A model-free information-theoretic measure of dependence between two random 
variables $ X $ and $ Y $ is the mutual information
\bea
\label{MI}
I(X;Y) &=& \mathbb{E}_{p(x,y)} \left[ \log \frac{p(X,Y)}{p(X) p(Y)} \right] =
\int dx dy \, p(x,y) \log \frac{p(x,y)}{p(x)p(y)} = 
\int dx dy \, p(x) p(y|x) \log \frac{p(y|x)}{p(y)} \nonumber \\
&=& \int dx dy \, p(x) p(y|x) \log \frac{p(y|x)}{\int p(x')p(y|x') dx'}
\eea
In particular, it follows by Jensen's inequality that $ I(X;Y) \geq 0 $, while the 
equality is reached iff the two variables are independent, i.e. $ p(x,y) = p(x) p(y) $,
see e.g. \cite{CT}.
It is easy to check that unconstrained minimization of (\ref{MI}) 
with respect to $ p(y|x) $ yields a factorized 
solution. Indeed, the variational derivative of (\ref{MI}) reads
\beq
\label{varder}
\frac{\delta I}{\delta p(y|x)} = p(x) \log \frac{p(y|x)}{p(y)} = 
p(x) \log \frac{p(x,y)}{p(x)p(y)}
\eeq
which vanishes when $ p(y|x) = p(y) $ or equivalently $ p(x,y) = p(x) p(y) $.

Let us now assume we are given constraints in the form of 
expectations $ \la F_{i}(x+y) \ra_{p(x,y)}  = C_i $. The minimum mutual information 
consistent with these constraints should be found by minimization of the following 
Lagrangian 
\beq
\label{LagrangeMI}
\mathcal{L} = I(X;Y) - \sum_{i} \lambda_i \left( \int dx d y \, p(x) p(y|x)
F_i (x + y) - C_i \right) - \int dx \, \xi(x) \left( \int dy \, p(y|x) - 1 \right)
\eeq
where $ \lambda_i $ and $ \xi(x) $ are Lagrange multipliers.
This variational problem has the formal solution
\beq
\label{formal}
p(y|x) = p(y) e^{ \sum_{i} \lambda_i F_i(x+y) + \xi(x)}
\eeq
While this solution is only a formal one as $ p(y) $ in turn depends on 
$ p(y|x) $ by the marginalization condition 
$ p(y) = \int dx \, p(x) 
p(y|x) $\footnote{If desired, the actual solution can be obtained using 
the Blahut-Arimoto alternating 
minimization algorithm, see \cite{CT}}, it
clearly shows that unless all $ \lambda_i = 0 $, 
the conditional probability $ p(y|x) $ does depend on $ x $ (as long as functions
$ F_i(x) $ are non-linear), i.e.
the independence solution $ p(x,y) = p(x) p(y) $ cannot hold anymore.
Physically, this means that the knowledge of expectations of functions of the 
form $ F_i (X+Y) $ carries information on interaction between two sub-portfolios
(i.e. has a non-vanishing information value), 
and is in general incompatible with the independence assumption\footnote{A similar 
observation was also made in \cite{GT} in the context of a related Minimum Information
principle for machine learning.}.    

 In our setting involving the market factor $ Z= \vec{Z}_{\vec{m}} $ described by 
a set of discretized probabilities $ \{ h_{\vec{m}} \} $, another relevant object is
the conditional mutual information of sub-portfolio losses $ X \equiv X_{11}$ and 
$ Y = X_{21} $ given $ Z_{\vec{m}} $:
\bea
\label{MIZ}
I(X;Y|Z) &=& \mathbb{E}_{p(x,y,z)} \left[ \log \frac{p(X,Y|Z)}{p(X|Z) p(Y|Z)} \right]
= \sum_{\vec{m}} h_{\vec{m}} \int dx dy \, P \left(x,y| \vec{m} \right)  
\log \frac{P(x,y| \vec{m})}{P_{X}(x| \vec{m}) P_{Y} (y| \vec{m})} \nonumber \\
&=& \sum_{\vec{m}} h_{\vec{m}} \int dx dy \, P_{X}(x| \vec{m}) P_{Y}(y|x, \vec{m}) 
\log \frac{P_{Y}(y|x, \vec{m})}{\int P_{X}(x'| \vec{m})P_{Y}(y|x', \vec{m}) dx' }
\eea
The same calculation with (\ref{MIZ}) results in factorization of 
conditional probabilities $ P(x,y | \vec{m}) = P_{X}(x | \vec{m}) P_{Y}(y| \vec{m}) $,
 i.e. in conditional independence.

Coming back to the unconstrained case, 
let us continue with the example of two random variables $ X $ and $ Y$ discussed 
above, which will now be identified with losses in sub-portfolios
$ \Pi_X $ and $\Pi_Y $, respectively. We assume that $ X $ and $ Y $ are 
independent {\it a priori}, which is the case when we condition on the market factor
$ Z $\footnote{Explicit dependence on $ Z $ is not important for the sake of discussion
in this section, and thus will be omitted below.}. 
As we have seen above, unconstrained minimization
of mutual information (MI) $ I(X;Y) $ yields the factorization (independence) 
result $ p(x,y) = 
p(x) p (y) $, but what should we take for marginal distributions $ p(x) $ and $ p(y) $?
Clearly, if we have some initial guesses (e.g. obtained with
historical calibration) $ q(x) $ and $q(y) $ for the marginals, it is natural to 
expect that in the absence of any constraints we should have $ p(x) = q(x) $, 
$ p(y) = q(y) $ and $ p(x,y) = p(x) p(y) = q(x) q(y) $. Formally, we can achieve this 
by minimization of the following functional
\beq
\label{Fun}
\mathcal{F}[p(x),p(y|x)] = I(X;Y) +  D[p(y)|| q(y)] +  D[p(x)||q(x)]
\eeq
where $ D[p||q] $ is the KL distance between two distributions $ p $ and $ q $.
Indeed, as all terms here are non-negative, the whole expression is minimized when each
term vanishes, which happens exactly for the case $ p(x) = q(x) $, 
$ p(y) = q(y) $ and $ p(x,y) = p(x) p(y) = q(x) q(y) $. We can now 
combine the first two 
terms in (\ref{Fun}) to get a more explicit expression
\beq
\label{Fun2}
\mathcal{F}[p(x),p(y|x)] = \int dx dy \, p(x) p(y|x) \log \frac{p(y|x)}{q(y)} + 
\int dx \, p(x) \log \frac{p(x)}{q(x)} = D[p(x,y)|| q(x) q(y)]
\eeq
In other words, we ended up with the objective function which is nothing but the 
KL distance between the unknown distribution $ p(x)$ and a factorized ``prior'' 
distribution $ q(x,y) = q(x) q(y) $. This provides a justification to the method 
used in the main text, where we mimimize KL distances between conditional
joint loss distributions of sub-portfolios and prior joint distributions, conditional
on pricing constraints that reference portfolios as a whole.

\jpmdisclaimer

\end{document}